\documentclass[twocolumn]{aastex631} 

\usepackage{tikz}
\usetikzlibrary{shapes.geometric, arrows}
\usetikzlibrary{positioning}

\tikzset{
    startstop/.style={rectangle, rounded corners, minimum width=2.6cm,
        minimum height=1cm, text centered, draw=black, fill=gray!20},
    process/.style={rectangle, minimum width=3.2cm, minimum height=1cm,
        text centered, draw=black, fill=blue!20},
    decision/.style={diamond, aspect=2, text centered,
        draw=black, fill=green!20, inner sep=1pt},
    arrow/.style={thick,->,>=stealth}
}

\begin{document}

\title{Formation Channels of Magnetars}

\author[0000-0002-6442-7850]{Rui-Chong Hu}
\affiliation{The Nevada Center for Astrophysics, University of Nevada, Las Vegas, NV 89154, USA}
\affiliation{Department of Physics and Astronomy, University of Nevada, Las Vegas, NV 89154, USA}
\email{ruichong.hu@unlv.edu}

\author[0000-0002-9725-2524]{Bing Zhang}
\affiliation{The Hong Kong Institute for Astronomy and Astrophysics, The University of Hong Kong, Pokfulam Road, Hong Kong, China}
\affiliation{Department of Physics, Department of Physics, The University of Hong Kong, Pokfulam Road, Hong Kong, China}
\affiliation{The Nevada Center for Astrophysics, University of Nevada, Las Vegas, NV 89154, USA}
\affiliation{Department of Physics and Astronomy, University of Nevada, Las Vegas, NV 89154, USA}
\email{bzhang1@hku.hk, bing.zhang@unlv.edu}

\begin{abstract}

The formation channels of magnetars remain an open question. Although core-collapse supernovae of isolated massive stars are important, binary interactions---such as tidal interaction, common envelope evolution, and stellar mergers---may also play a significant role in making magnetars. Understanding the relative contributions of these channels is crucial for linking magnetars to their observed properties and host environments. In this paper, we investigate potential magnetar formation channels using population synthesis simulations, considering both single-star and isolated binary system evolution. By conducting simulations with different parameters, we compare the effects of various evolution processes on magnetar formation. Additionally, we study the delay times and kick velocities across all formation channels, analyze the orbital properties and companion types of surviving magnetar binaries. We find that the majority of magnetars are observed as single objects ($\geq 90\%$), although a large fraction of them were originally in binary systems and experienced either kick disruption or merger. Surviving binaries are most likely to host main-sequence companions and exhibit different distributions of eccentricities due to different supernova mechanisms. These findings show the critical role of binary evolution in magnetar formation and provide predictions for the properties of magnetar populations that can be tested with future observations.

\end{abstract}


\section{Introduction} \label{sec:intro}

Magnetars are a type of young neutron star (NS) characterized by their extreme magnetic fields, observed across X-ray, gamma-ray, and radio bands \citep[e.g.,][]{Turolla2015,Kaspi2017}.
To date, observations have identified 29 magnetars in the Milky Way and the Magellanic Clouds \citep{Rea2025}, with most detected as soft gamma repeaters (SGRs) and anomalous X-ray pulsars (AXPs).
These sources exhibit bursts of high-energy radiation, which is believed to be powered by their strong magnetic fields \citep{Thompson1995,Thompson1996}. Furthermore, magnetars have been proposed as the likely sources of cosmological fast radio bursts (FRBs) \citep{lorimer2007,zhang2023}, with observational evidence linking one FRB (FRB 20200428D) to a known Galactic magnetar (SGR J1935+2154) \citep{CHIME-SGR,STARE2-SGR,Integral-SGR,HXMT-SGR,Tavani2021,Ridnaia2021}.

Magnetars are also believed to be the central engine powering gamma-ray bursts \citep[GRBs;][]{Usov1992,Thompson1994,Dai1998,Zhang2001,Metzger2011}, superluminous supernovae \citep[SLSNe;][]{Kasen2010,Woosley2010,Inserra2013}, broad-line Type Ic supernovae \citep[SNe Ic-BL;][]{Yu2017,Zhangbook,Zhu2025}, and fast blue optical transients \citep[FBOTs;][]{Yu2015,Rest2018,Margutti2019}, where the spin-down of the newly formed magnetar injects additional energy into the ejecta, significantly enhancing their luminosity. 
 
Two main classes of mechanisms have been proposed to account for the extreme magnetic fields of magnetars. One is that magnetic fields may originate as fossil remnants of their progenitors \citep{Braithwaite2004,Ferrario2006}. 
The other class involves magnetic-field amplification via dynamo processes. These include the convective dynamo, which operates through turbulent convection in a rapidly rotating proto-NS \citep[PNS;][]{Thompson1993,Raynaud2020,Masada2022}; the magnetorotational-instability-driven dynamo, which operates in rapidly rotating PNSs where shear instabilities amplify seed magnetic fields \citep{Obergaulinger2009,Mosta2014,Reboul2021}; and the Tayler-Spruit dynamo, which is driven by differential rotation and Tayler instability due to fallback accretion onto the PNS \citep{Barrere2022,Barrere2023,Barrere2025}.

Those mechanisms are closely related to the evolutionary history of magnetar progenitors and the magnetar physical conditions at birth. In the fossil field scenario, it remains uncertain whether a sufficient fraction of massive stars host magnetic fields of the necessary strength to account for the magnetar-level field \citep{Makarenko2021}.
Alternatively, dynamo-based amplification requires the PNS to retain significant angular momentum (AM), either inherited from a rapidly rotating progenitor core or gained through fallback accretion during the core-collapse supernova (CCSN). This raises two key questions: which evolution channels can produce a rapidly rotating stellar core and under what birth conditions can sufficient fallback accretion occur to trigger the Tayler-Spruit dynamo.

Observational studies have shown that a significant portion of massive stars reside in binary or higher-order multiple systems \citep{Sana2012,Sana2014}. These suggest that binary interactions may play a significant role in magnetar formation \citep[e.g.,][]{Popov2006,Popov2016,Popov2023}. In close binaries, tidal interactions can spin up progenitor stars (helium stars), enabling magnetic-field amplification and potentially leading to magnetar formation \citep{Bogomazov2009,Fuller2022,Hu2023}. Another proposed channel is the core merger-induced collapse (CMIC), in which a magnetized oxygen–neon–magnesium (ONeMg) white dwarf (WD) merges with a nondegenerate stellar core during the common envelope (CE) phase, resulting in the collapse to a NS with a strong magnetic field \citep{Ablimit2022b}. Additionally, mergers of compact objects, such as binary WDs \citep[BWDs;][]{Schwab2021}, WDNS systems \citep{Zhong2020}, or binary NSs \citep[BNSs;][]{Giacomazzo2013,Reboul-Salze2024}, have also been proposed as potential channels for forming magnetars under certain conditions. 
Understanding these different formation channels is crucial for linking magnetars to their observational counterparts in different astrophysical environments.

In this study, we investigate both single-star evolution and isolated binary evolution channels of magnetar formation using population synthesis simulations. In section \ref{sec:method}, we introduce the method of population synthesis simulation and the origins of magnetar magnetic fields. In section \ref{sec:results}, we present the main results and compare them with observations. In section \ref{sec:discussion}, we discuss the implications for magnetar-related transients and main uncertainties in our simulation. In section \ref{sec:conclusion}, we summarize our main conclusions.

\section{Method} \label{sec:method}

\subsection{Population Synthesis Simulations} \label{sec:PS method}

We use the rapid binary population synthesis code \texttt{COMPAS}\footnote{https://github.com/TeamCOMPAS/COMPAS} {\citep[version v03.12.00;][]{Stevenson2017,Vigna2018,Neijssel2019,TeamCOMPAS2022,TeamCOMPAS2025}} to explore magnetar formation through both single-star evolution (``SSE'' mode) and isolated binary-star evolution (``BSE'' mode) channels. 
Our fiducial population synthesis model is based on the setup described in Table 1 of \cite{Broekgaarden2021}, which assumes an initial mass function (IMF) from \cite{Kroupa2001}, a flat mass ratio distribution and a log-uniform separation distribution, and includes prescriptions for wind mass loss \citep{Belczynski2010a,Belczynski2010b}, CE evolution \citep[$\alpha-\lambda$ formalism;][]{Webbink1984,DeKool1990} and supernova (SN) remnant formation, following \cite{Fryer2012}. 
Although these fiducial assumptions provide a well-tested baseline, a number of key evolutionary processes remain uncertain. To assess their impact, we systematically vary several parameters and physics prescriptions across different model configurations, simulating $10^6$ systems for each single and binary model. The full list of parameters is provided in Appendix \ref{sec: app_setup}.

The stellar winds of massive stars, such as helium stars, remain an open question in stellar evolution \citep{Vink2022}. 
In our fiducial model, we use metallicity-dependent stellar wind prescriptions, following the settings in \cite{vanSon2025} and \cite{Merritt2025}, based on the latest research development. 

During the mass transfer phase, the fraction of mass lost by the donor that is accreted by the accretor is parameterized by a factor $\beta$, such that $\dot{M}_{\text{acc}} = -\beta \dot{M}_{\text{don}}$, where $0 \leq \beta \leq 1$. In our fiducial model, if the donor transfers more mass than the accretor can accept, the excess is assumed to be lost from the vicinity of the accreting star via isotropic reemission \citep[e.g.][]{Bhattacharya1991,Tauris2006}. For comparison, we also investigate a fixed accretion efficiency of $\beta = 0.5$.

CE evolution is a key phase in the formation of close compact binaries, where unstable mass transfer leads to the engulfment of the companion in the donor envelope. In the $\alpha-\lambda$ formalism \citep{Webbink1984,DeKool1990}, the parameter $\alpha$ measures how efficiently the orbital energy is used to eject the envelope, while $\lambda$ describes its binding energy. In our simulation, we choose $\alpha_{\mathrm{CE}}=1$ in our fiducial model, and explore the variations of different CE efficiencies ($\alpha_{\mathrm{CE}} = 0.5,\,2,\,5$).
 
Since the LIGO–Virgo–KAGRA (LVK) collaboration reported the mass-gap black hole (BH)–NS merger event GW230529 \citep{LVK2024}, we adopt the ``delayed'' SN prescription \citep{Fryer2012} to determine compact object masses during CCSNe in our fiducial model, which allows for the production of mass-gap BHs.
Due to the asymmetry of the ejecta, the remnant typically receives a natal kick \citep[e.g.][]{Janka1994,Wongwathanarat2013,Popov2025}, which we model with a Maxwellian distribution with dispersions of $\sigma_{\mathrm{CCSN}}=265\,{\rm km}\,{\rm s}^{-1}$ \citep{Hobbs2005} in our fiducial model and $\sigma_{\mathrm{CCSN}}=100\,{\rm km}\,{\rm s}^{-1}$ for comparison. We also consider the recent model proposed by \cite{Disberg2025}, which adopts a lognormal distribution with $\mu = 5.6$ and $\sigma = 0.68$. For electron-capture SNe (ECSNe) and ultrastripped SNe (USSNe), we adopt weaker natal kicks, as their ejecta masses are believed to be lower than those of CCSNe \citep{Pfahl2002,Podsiadlowski2004}.

We adopt a maximum NS mass of $2.5 M_{\odot}$ in our fiducial model, and we also consider $2.2 M_{\odot}$ for comparison.
Given the significant uncertainties in SN modeling---and the impact of fallback accretion onto the PNS on magnetar formation, as introduced in \ref{sec:magnetic field}---we compare the outcomes using the ``rapid'' SN prescription \citep{Fryer2012} and the ``startrack'' prescription \citep{Belczynski2002} in section \ref{sec:discussion}.

To compare the contributions from single-star evolution and isolated binary channels, it is important to account for the fraction of massive stars that can produce NSs in binary systems. Observational studies have shown that a significant portion of massive stars are in binary or higher-order multiple systems. \cite{Sana2012} found that more than 70\% of O stars are not effectively single at birth and \cite{Moe2017} found that approximately 84\% of early B-type stars are in binary, tertiary, or higher-order systems. Considering both constraints, we assume that $80\%$ massive stars form in binary systems in our simulation.

Since the typical lifetime of a magnetar is less than $10^6$ yr, which is much shorter than the evolutionary timescale of galaxies, it is reasonable to use a fixed metallicity $Z=Z_{\odot} = 0.0142$ \citep{Asplund2009} for simulations of the Galactic magnetar population.
We refer the reader to Table \ref{tab:COMPAS} in Appendix \ref{sec: app_setup} for further details on other initial conditions, parameter settings, and simulation settings we used.

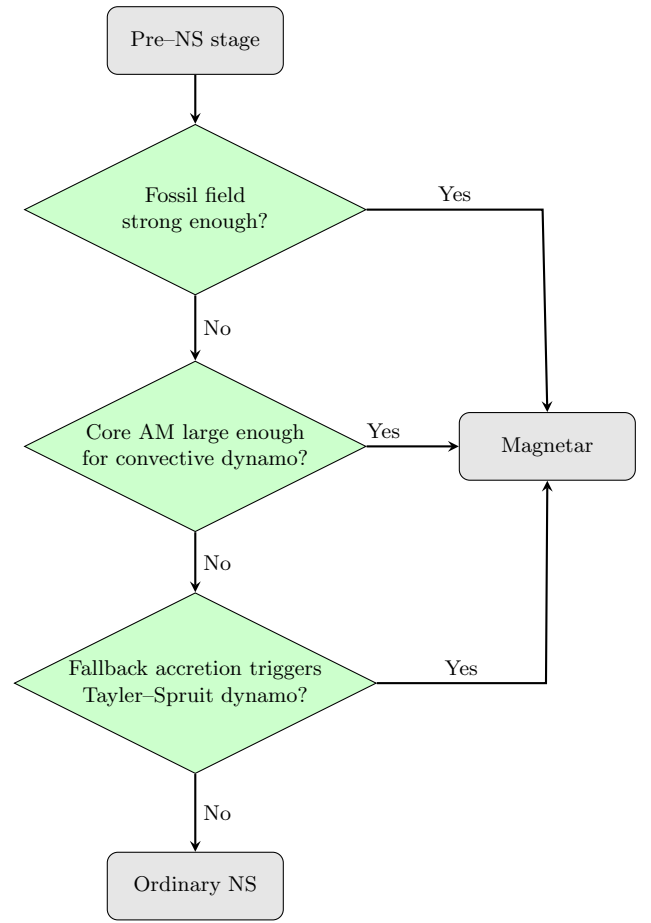
\begin{figure}[ht!]
\begin{center}
    \begin{tikzpicture}[node distance=2cm, scale=0.9, every node/.style={transform shape}]

    \node (start) [startstop] {Pre--NS stage};
    \node (fossil) [decision, below of=start, yshift=-0.5cm] 
        {\parbox{3.5cm}{\centering Fossil field\\ strong enough?}};
    \node (convective) [decision, below of=fossil, yshift=-1.5cm] 
        {\parbox{3.5cm}{\centering Core AM large enough\\ for convective dynamo?}};
    \node (tayler) [decision, below of=convective, yshift=-1.5cm] 
        {\parbox{3.8cm}{\centering Fallback accretion triggers\\ Tayler--Spruit dynamo?}};
    
    \node (mag) [startstop, right of=convective, xshift=3.2cm] {Magnetar};
    \node (ordinary) [startstop, below of=tayler, yshift=-1.0cm] {Ordinary NS};

    \draw [arrow] (start) -- (fossil);

    \draw [arrow] (fossil.east) -- ++(2.6cm,0) node[midway,above]{Yes} -- (mag.north);
    \draw [arrow] (fossil.south) -- (convective.north) node[midway,right]{No};

    \draw [arrow] (convective.east) -- ++(0.5cm,0) node[midway,above]{Yes} -- (mag.west);
    \draw [arrow] (convective.south) -- (tayler.north) node[midway,right]{No};

    \draw [arrow] (tayler.east) -- ++(2.5cm,0) node[midway,above]{Yes} -- (mag.south);
    \draw [arrow] (tayler.south) -- (ordinary.north) node[midway,right]{No};
    \end{tikzpicture}
\end{center}
    \caption{Flowchart illustrating the decision process for magnetar formation. A PNS becomes a magnetar if any of the following conditions are satisfied: (1) a sufficiently strong fossil magnetic field is inherited from the progenitor, (2) rapid core rotation provides enough AM and differential rotation to drive a convective dynamo, or (3) fallback accretion during core collapse triggers the Tayler–Spruit dynamo. If none of these amplification mechanisms operate, the outcome is an ordinary NS.}
    \label{fig:flowchart}
\end{figure}

\subsection{Origin of Strong Magnetic Fields} \label{sec:magnetic field}

The strong magnetic field is the key property we use to distinguish magnetars from normal NSs, so the origin of such fields becomes very important. Figure \ref{fig:flowchart} provides a summary of how we determine whether the end product is a magnetar or a normal NS.
The first mechanism we considered is the fossil field hypothesis \citep{Braithwaite2004,Ferrario2006}, which suggests that magnetars inherit their strong magnetic fields from their progenitor stars.  

The strong magnetic field of a progenitor main-sequence (MS) star could originate from the merger of two MS stars (we have not considered MS stars with an initially strong magnetic field here, as the mechanism by which a massive star can sustain such a field remains under debate; e.g., see \citealp{Frost2024}). Magnetohydrodynamical simulations showed that the merger of two MS stars can generate a rapidly rotating massive star with strong magnetic fields \citep{Schneider2019,Vynatheya2025}, eventually collapsing to form a magnetar. Since only about 7\% of massive stars are observed to display large-scale magnetic fields of hundreds to thousands of Gauss  (e.g. \citealt{Frost2024}, and references therein), and this fraction is similar to the fraction of massive stars experiencing a merger in our simulation (see section \ref{sec:results}), we assume that all mergers of massive stars that ultimately produce NSs result in magnetars. For comparison, we also consider the case in which massive star binary mergers produce magnetars only through the activation of the Tayler–Spruit dynamo (see section \ref{sec: uncertainties_BMS}).

The merger of two WDs, the merger of an ONeMg WD with a nondegenerate core, and the accretion-induced collapse (AIC) of an ONeMg WD could also generate a magnetar, if the WD possesses a strong magnetic field. Here, we only consider the cases where the total mass of the two WDs exceeds 1.4 $M_{\odot}$, and the parameter space for AIC is consistent with \cite{Ablimit2022a}, which results in NS formation. And we consider 15\% of these WDs possess strong magnetic fields \citep{Ferrario2015}, which can lead to the formation of magnetars.

In addition to the fossil field hypothesis, dynamo mechanisms could also amplify magnetic fields and result in magnetar formation. Helium stars in close binaries can be spun up through strong tidal interactions with their companions, leading to rapid rotation \citep[e.g.,][]{vandenHeuve2007,Qin2018}. Differential rotation can amplify magnetic fields through the Tayler–Spruit dynamo \citep{Spruit2002}, and the fast-spinning core of a PNS may trigger the convective dynamo at the early stage to generate a strong magnetic field. Such a channel may occur after a CE phase or stable mass transfer (SMT), where the helium core is left in a highly spun-up state before collapsing \citep{Fuller2022,Hu2023}. Here we adopt the parameter space of \cite{Hu2023} to identify helium stars in binary systems as the potential progenitor of magnetars. 

Another possibility is that, during CCSNe, a fraction of the ejected material may fall back onto the PNS, leading to an accretion-driven spin-up. If the accreted mass is sufficiently large (typically 
$10^{-2} M_{\odot}$ to $10^{-1} M_{\odot}$), the resulting differential rotation can trigger the Tayler-Spruit dynamo and amplify the magnetic field \citep{Barrere2022}. 
Since the fallback accretion mass depends on the uncertain SN mechanism, we apply different magnetar formation fractions among all NSs to constrain the fraction of magnetars formed through CCSNe. Based on observations and previous studies, we adopt 50\% as our fiducial value \citep{Pardo2026} and compare the results with those obtained using 80\% and 20\%.

\section{Results} \label{sec:results}

\begin{figure*}[p]
\centering
\includegraphics[width=0.80\linewidth, trim = 0 0 0 0, clip]{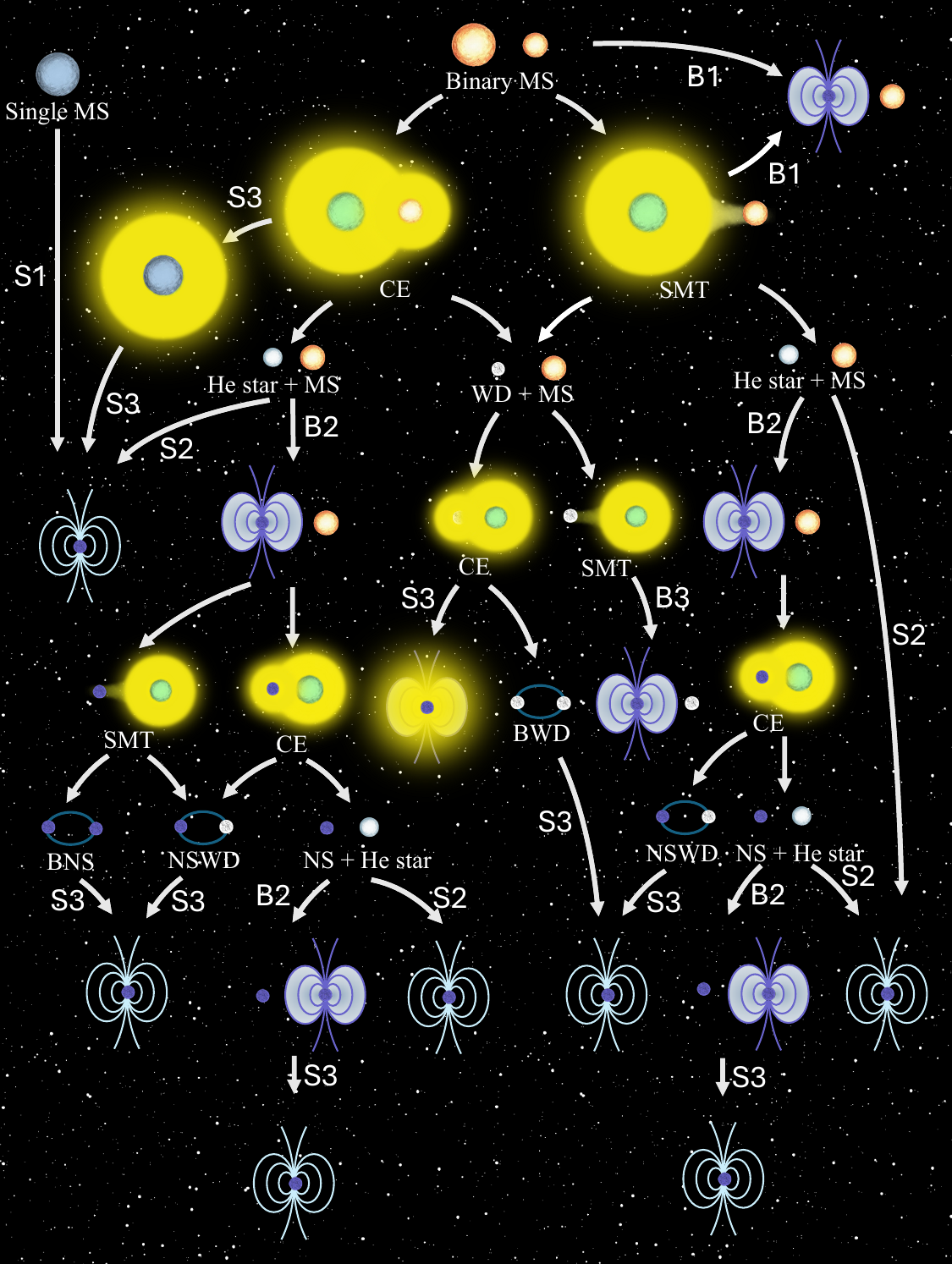}
\caption{Flowchart of formation of magnetars through single and isolated binary evolution channels. Magnetars with a light-colored magnetic-field background are in binary systems, while those with a black-colored background are single.}
\label{fig:illustration}
\end{figure*}

In this section, we present the results of our population synthesis simulations. As shown in Fig.\ref{fig:illustration}, magnetar formation channels can be categorized into two types: those in which magnetars form as single objects (“S”) and those in which they form within binary systems (“B”). The numbers indicate different subchannels within each category. The fractions of different channels presented in sub-sections \ref{sec:single} and \ref{sec:binary} are based on our fiducial population synthesis model.

\subsection{Channels forming single magnetars}\label{sec:single}

\subsubsection{Single-Star Evolution: S1 Channel}
    In our simulations, a magnetar could form through the evolution of a single massive star. During core collapse, the PNS is spun up by fallback accretion. The Tayler-Spruit dynamo then amplifies the large-scale magnetic field, leading to the formation of a magnetar \citep{Barrere2022,Barrere2023,Barrere2025}. 
    This channel contributes approximately 19\% of the total magnetar population. The initial masses of the progenitors in this channel range from about 5 $M_\odot$ to 21 $M_\odot$.
\subsubsection{Binary Disruption: S2 Channel}
    A large fraction of massive star binary systems are disrupted by natal kicks caused by asymmetric SN explosions \citep[e.g.,][]{Brandt1995}. If a magnetar already exists in the system or is formed during the explosion, the disruption of the binary will result in an isolated magnetar \citep{Sherman2024}. Prior binary interactions may have enhanced the rotation of the progenitor star, triggering a dynamo mechanism that amplifies the magnetic field and ultimately leads to the formation of a magnetar. Alternatively, the magnetar could form through a CCSN that triggers the Tayler-Spruit dynamo. This channel contributes approximately 57\% of the total magnetar population and becomes the dominant magnetar formation channel.
\subsubsection{Stellar Mergers: S3 Channel}
    If the initial separation of two MS stars is small enough or the separation is decreased during the evolution of the binary system, the two MS stars may merge directly or form a CE, which could also lead to a merger if the system fails to eject the CE. Such a merger may produce a massive star with a strong, large-scale surface magnetic field and could ultimately evolve into a magnetar \citep{Schneider2019,Shenar2023,Frost2024,Vynatheya2025}.
    Based on our fiducial model, mergers of MS stars that produce remnants massive enough to finally collapse into NSs account for roughly 8\% of all massive stars, and this channel contributes to about 19\% of the total magnetar population. The CCSNe that form magnetars following stellar mergers can contribute to ``late'' massive star explosion events \citep{Zapartas2017}.
    
    If the merger occurs during the CE phase of a magnetized ONeMg WD and the core of a hydrogen-rich or helium-rich nondegenerate star, it can result in the formation of a magnetar surrounded by a massive envelope \citep{Ablimit2022b}. This channel contributes only about 0.1\% of the total magnetar population. 
    
    Magnetar can form through compact object mergers, including BWD mergers \citep{Schwab2021}, WD-NS mergers \citep{Zhong2020}, and BNS mergers \citep{Reboul-Salze2024}. Since it takes a long time for gravitational wave (GW) radiation to shrink the orbit and drive the binary toward merger, magnetars formed through this channel have a relatively long delay time compared to those formed through CCSNe
    and are relevant for old stellar populations.
    Since the merger rate of BNSs is low \citep[e.g.,][]{LVK2025}, and only a fraction of such mergers are expected to produce stable, long-lived magnetars (most of them would collapse within seconds to a year after the mergers; e,g, \citealt{ai2020}), we do not include their contribution in our analysis.

\subsection{Channels forming magnetars in binaries}\label{sec:binary}

\subsubsection{Binaries Survived through Supernovae: B1 Channel}
    In some binary systems where the magnetic field of the progenitor star is not significantly amplified prior to core collapse, it is still possible for a magnetar to form during the CCSN by triggering the Tayler-Spruit dynamo.
    If the magnitude and direction of the natal kick happen to allow the binary system to survive the SN explosion, it then results in a magnetar that remains bound within a binary system.
    This channel contributes to about 1\% of the total magnetar population and 25\% of the binary magnetar population.
\subsubsection{Tidal Spin-Up: B2 Channel}
    When a star in a binary system eventually expands and fills its Roche lobe, it may initiate a mass transfer phase onto its companion.
    If the mass transfer remains stable and successfully removes the donor’s entire hydrogen envelope, the star becomes a stripped-envelope, helium-burning object, commonly referred to as a helium star. On the other hand, if the mass transfer is dynamically unstable, it leads to a CE phase. During this phase, the orbital separation of the binary shrinks as the orbital energy and AM are transferred to the envelope \citep{Ivanova2013}. If the CE is successfully ejected, the system emerges as a close binary consisting of a massive helium star and its companion. Strong tidal interactions between the two stars can spin up the helium star. 
    This enhanced rotation leads to the formation of fast-spinning magnetars at the end of their lives \citep{Fuller2022,Hu2023}. Most magnetar binaries originate from this channel ($\sim 75\%$), which account for about 3\% of the total magnetar population.
\subsubsection{Accretion-Induced Collapse: B3 Channel}
    In this channel, an ONeMg WD in a close binary system accretes mass from a companion star. Once the WD approaches the Chandrasekhar mass limit, it undergoes an AIC, forming a NS. If the progenitor WD possesses a strong magnetic field, the resulting NS may be born as a magnetar \citep{Ablimit2022a}. This channel contributes only roughly 0.1\% of the total magnetar population.
    
\subsection{Fractional estimates}
\label{sec:fractions}

There are great uncertainties in the fraction of magnetars born in all CCSNe. Theoretically, the amount of fallback accretion mass is rather uncertain, which limited the predictive power of the fraction of CCSNe where Tayler-Spruit dynamo operates. Observationally, there is evidence that this fraction is close to 1/2, i.e. the ratio between magnetars to normal-field NSs is about 1:1 \citep{Pardo2026}. Hereafter we will adopt this ratio in our fiducial model, but discuss the effect of this ratio later (see section \ref{sec: uncertainties_CCSN}). 

\begin{figure*}
\centering
\includegraphics[width=0.75\linewidth, trim = 0 0 0 0, clip]{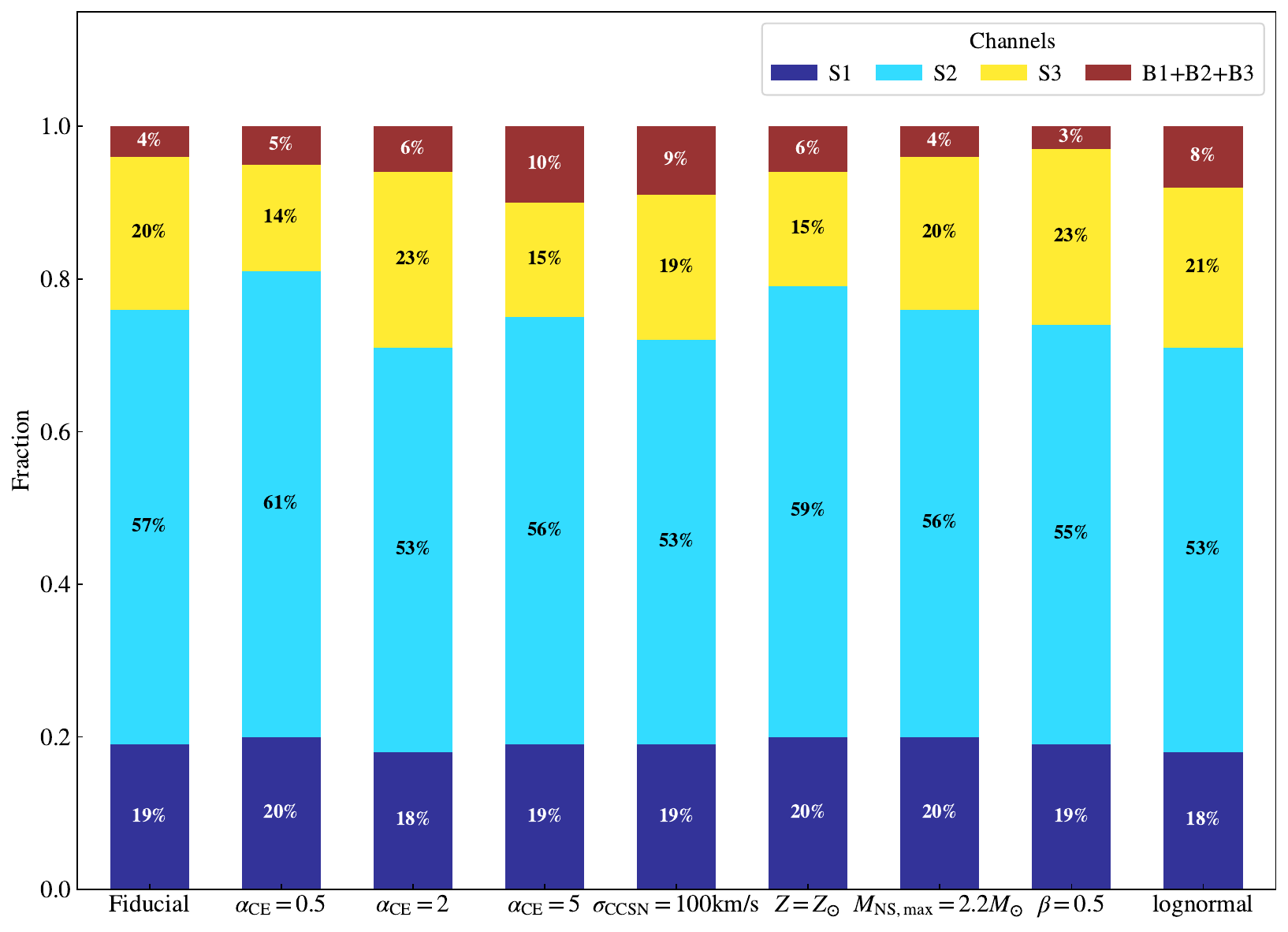}
\caption{Fractions of different magnetar formation channels based on different population synthesis models.}
\label{fig:channel_fractions}
\end{figure*}

Based on the conditions of different magnetar formation channels discussed above, we calculate the relative fractions contributed by each channel to the overall magnetar population and present them in Fig.\ref{fig:channel_fractions}. We find that, across all channels, the majority of the magnetars are likely observed as isolated objects ($\geq 90\%$). Additionally, most magnetars are produced through CCSNe.

These results also suggest that the S2 channel is the dominant channel across all channels, which means that even if most magnetars appear isolated, a large fraction may have undergone significant binary interaction prior to formation. This has implications for interpreting magnetar environments and finding potential companions of magnetars. Observationally, searches for stellar companions to Galactic magnetars have placed stringent upper limits on the binary magnetar fraction, implying that the observable population of binary magnetars is very small \citep[$5\sim10\%$;][]{Chrimes2022}. This is consistent with our results.

Compared to the fiducial model, adopting a higher CE efficiency ($\alpha_{\mathrm{CE}} > 1$) leads to tighter post-CE binaries, hence increasing the fraction of magnetars that remain in binary systems. Models with reduced natal kicks ($\sigma_{\mathrm{CCSN}}=100$km/s or lognormal model) also result in a higher fraction of magnetars in bound binaries and fewer in the SN-disrupted category, as more systems survive the SN explosions. If we assume a fixed mass transfer efficiency of $\beta = 0.5$, the contribution from binary channels becomes smaller since more binaries will merge before forming an NS.

\subsection{Properties of Magnetars from Different Channels}
\label{sec:properties}

\subsubsection{Delay Time Distribution}

\begin{figure*}
\centering
\includegraphics[width=1.0\linewidth, trim = 0 0 0 0, clip]{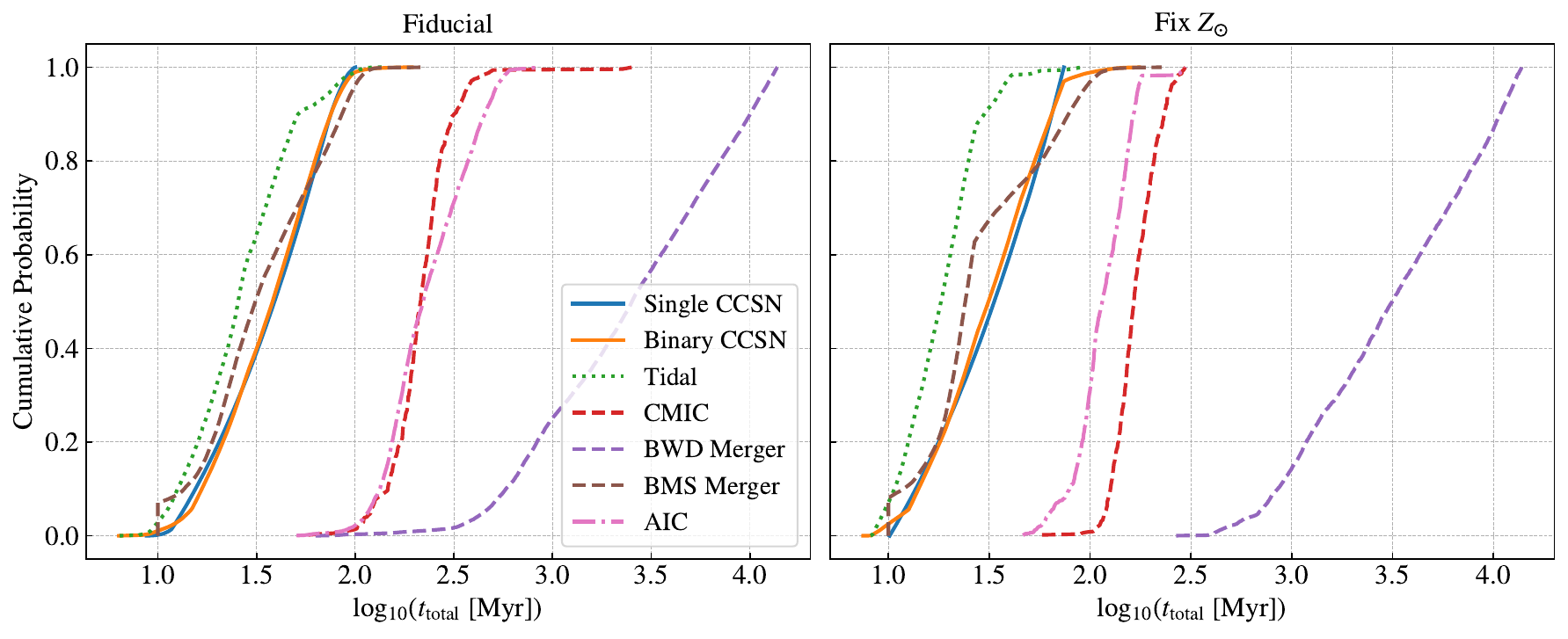}
\caption{Cumulative distribution functions of delay times for different evolutionary channels. The left panel shows the results for the fiducial model, while the right panel presents a model with metallicity fixed as $Z_{\odot}$. Each curve represents a different formation scenario, including single and binary CCSNe, tidal interactions, mergers (BWD and binary MS (BMS)), and AIC. The x axis denotes the logarithm of the total delay time in Myr, while the y axis shows the cumulative probability.}
\label{fig:delay_time}
\end{figure*}

Figure \ref{fig:delay_time} presents the cumulative delay time distributions of magnetars formed through different channels, compared to the star formation history. This delay time highlights how quickly magnetars are expected to form after the progenitor formation, across different channels.
The left panel shows results from the fiducial model, which includes metallicity evolution, while the right panel shows results assuming a constant solar metallicity ($Z=Z_{\odot}$).

\begin{figure*}
\centering
\includegraphics[width=1.0\linewidth, trim = 0 0 0 0, clip]{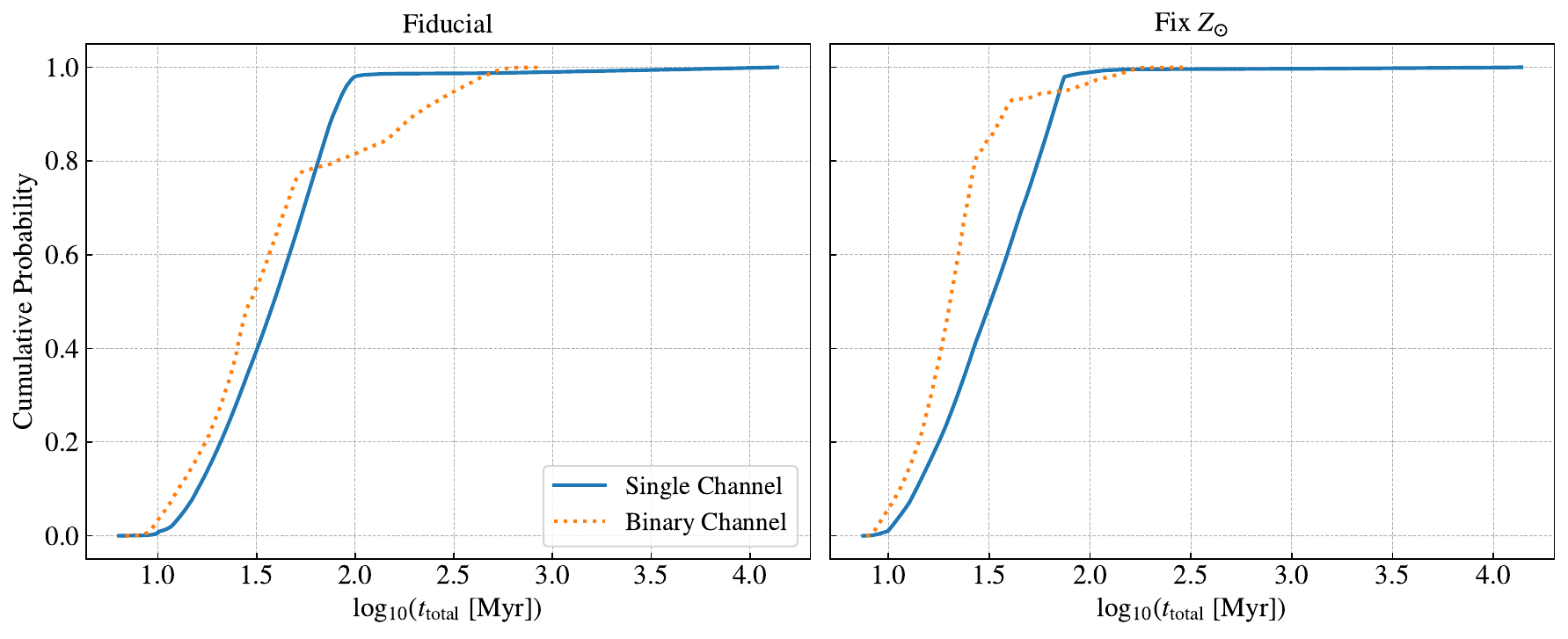}
\caption{Same as Fig.\ref{fig:delay_time}, but showing the distribution from the total single and total binary channels.}
\label{fig:delay_time_S_B}
\end{figure*}

In both models, the delay-time distributions can be broadly grouped into three categories. The BWD merger channel exhibits the longest delay times, as the mergers are driven by GW radiation, which operates over very long timescales. The CMIC and AIC channels show intermediate delay times, since both involve the formation and evolution of ONeMg WDs—processes that require extended evolutionary times compared to the evolution of massive stars but are still shorter than the merger timescales of compact objects. The remaining channels have relatively short delay times, comparable to those of the single CCSN channel, as they are primarily governed by the lifespans of massive stars.

The fixed-metallicity model produces faster accumulation of magnetars across nearly all channels except for the BWD merger channel. This indicates systematically shorter delay times at fixed solar metallicity. This difference arises because higher-metallicity stars ($Z=Z_{\odot}$) typically evolve faster due to stronger stellar winds and shorter lifespans. As a result, channels that depend on massive star evolution lead to earlier magnetar formation.
In contrast, the WD merger channel shows a slightly longer delay time. This is expected, as this channel depends primarily on GW radiation timescales, which are relatively insensitive to progenitor metallicity. 

To further clarify the role of the different evolutionary pathways, we group the delay time distributions into two broad categories: single and binary channels. 
As shown in Fig.\ref{fig:delay_time_S_B}, in general, single channels tend to contribute to the prompt population, producing magnetars on short timescales. Binary channels broaden the distribution substantially: very short delays arise from rapid tidal spin-up, while much longer delays appear through channels such as AIC. The absolute longest delay times in our simulations are associated with BWD mergers, which we classify as single channel. However, these events are very rare and only contribute a small tail to the overall distribution. This difference highlights the potential of delay time distribution studies to distinguish single and binary contributions.

\subsubsection{Companion Types and Masses}

\begin{figure*}
\centering
\includegraphics[width=0.75\linewidth, trim = 0 0 0 0, clip]{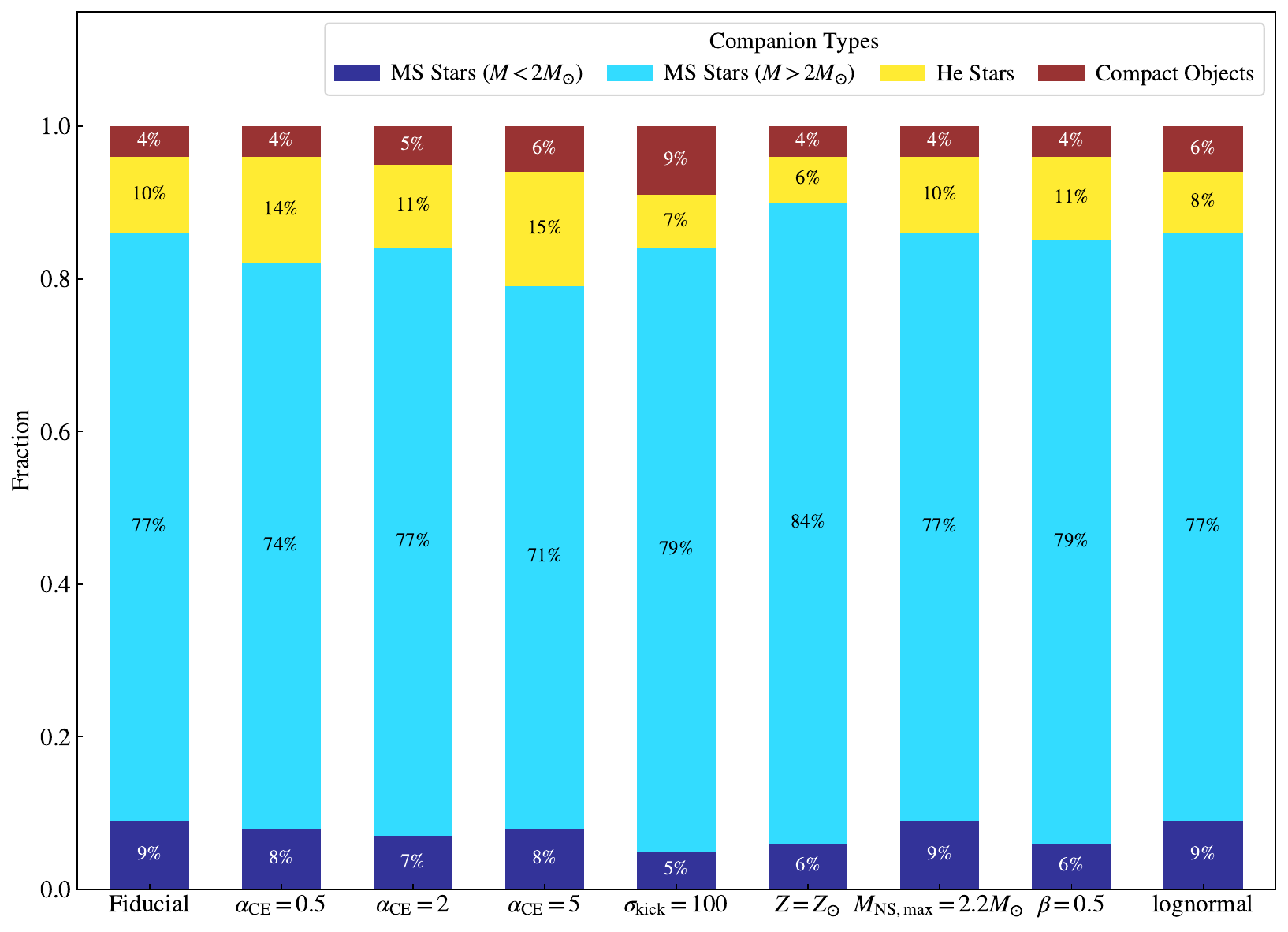}
\caption{Fractions of different magnetar companion types based on different population synthesis models.}
\label{fig:companion_fractions}
\end{figure*}

\begin{figure*}
    \centering
    \includegraphics[width=0.49\textwidth]{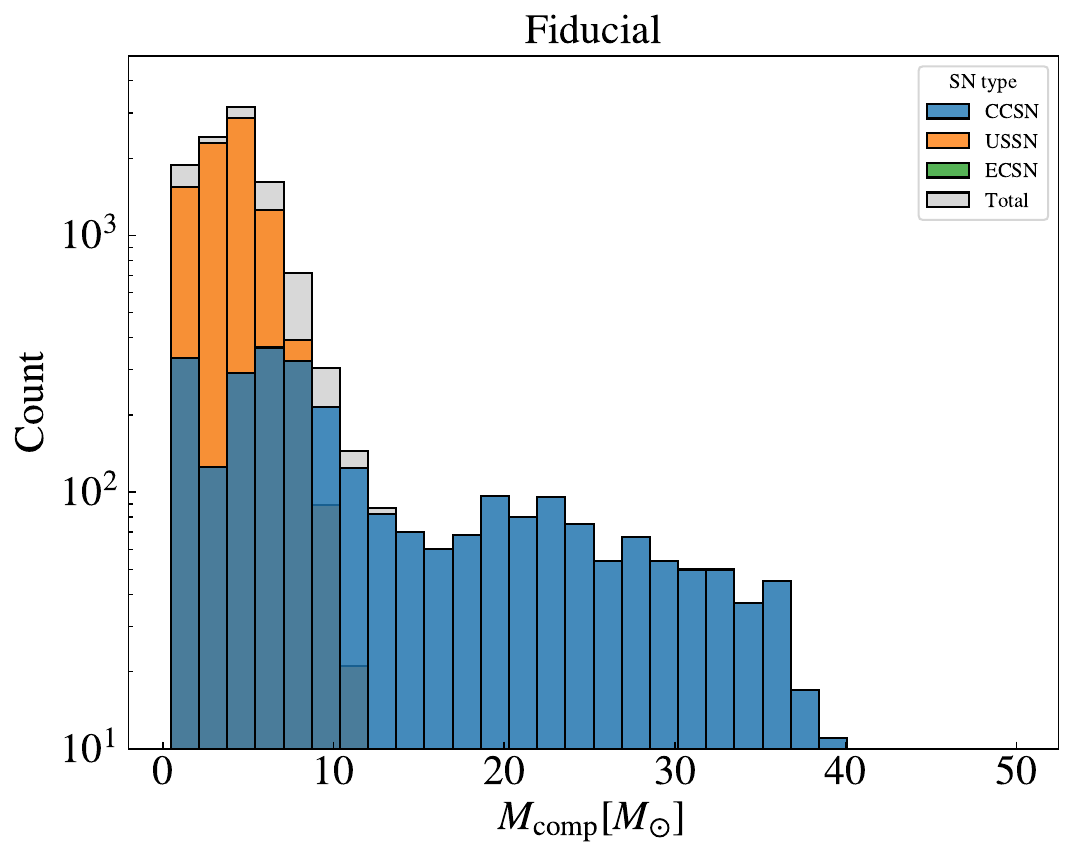}\hfill
    \includegraphics[width=0.49\textwidth]{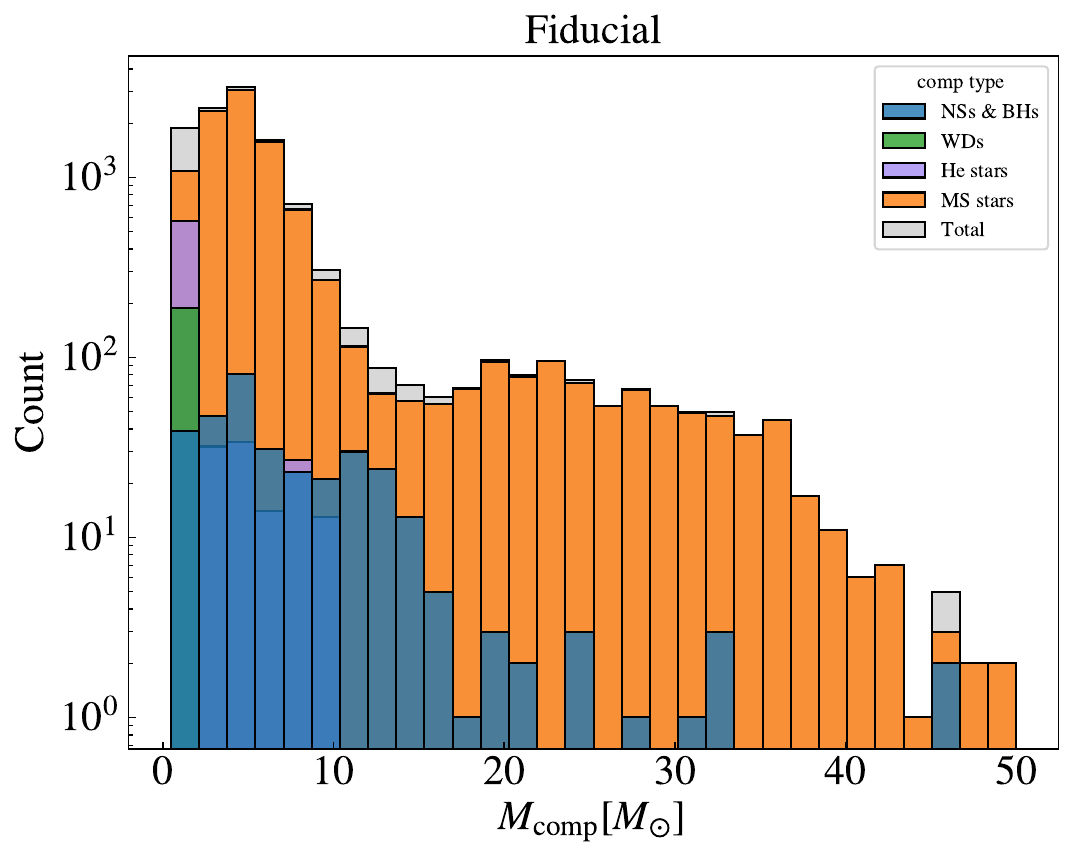}\hfill
    \caption{Distribution of magnetar companion masses. Left panel: different colors represent different SN mechanisms, while the gray bars indicate the total distribution. Right panel: different colors represent different companion types.}
    \label{fig:comp_mass}
\end{figure*}

Figure \ref{fig:companion_fractions} presents the distribution of magnetar companion types across different formation channels. 
Companions are categorized into four types: massive MS stars ($M>2M_{\odot}$), low-mass MS stars ($M<2M_{\odot}$), helium stars, and compact objects (WDs, NSs or BHs).
Across all models, MS stars are the dominant companion type, which consistently contributes 79–90\% of the total (for the solar metallicity model, it will increase to 90\%). Helium star companions appear in a smaller fraction (6–15\%), while compact object companions are relatively rare, typically below 10\%.

These results suggest that, if magnetars remain in binary systems, their companions are most likely to be MS stars with masses exceeding $2M_{\odot}$, potentially corresponding to OB type stars (see also \citealt{zhanggao2020}). For instance, certain FRB scenarios invoke a magnetar-OB star binary, where the stellar wind and magnetized environment play key roles in shaping observable properties such as rotation measure (RM) variation and potential RM sign reversal. The presence of such companions in our population synthesis results supports the possibility that these systems could be the sources of actively repeating FRBs \citep{Zhang2025}. A small fraction of surviving magnetar binaries host low-mass helium star companions, with masses down to about $0.8 M_{\odot}$. These systems typically experienced a CE or SMT phase, and the companion may become a subdwarf-B star \citep{Wu2018}.

The distribution of companion masses for magnetars in the fiducial model is shown in the left panel of Fig.\ref{fig:comp_mass}. The contribution from ECSNe is negligible compared to the other two channels. 
USSNe dominate among the surviving binaries, as they are predicted to produce weaker natal kicks than CCSNe due to their lower ejecta masses. This results in a higher probability of the binary system remaining bound after the explosion. In contrast, the higher kicks associated with CCSNe disrupt most binaries, resulting in the formation of more isolated magnetars. As shown in the right panel of Fig.\ref{fig:comp_mass}, the majority of companions in magnetar binaries have masses between 2 and 10 $M_{\odot}$, while MS stars dominate the high-mass end of the distribution.

\begin{figure*}
\centering
\includegraphics[width=0.75\linewidth, trim = 0 0 0 0, clip]{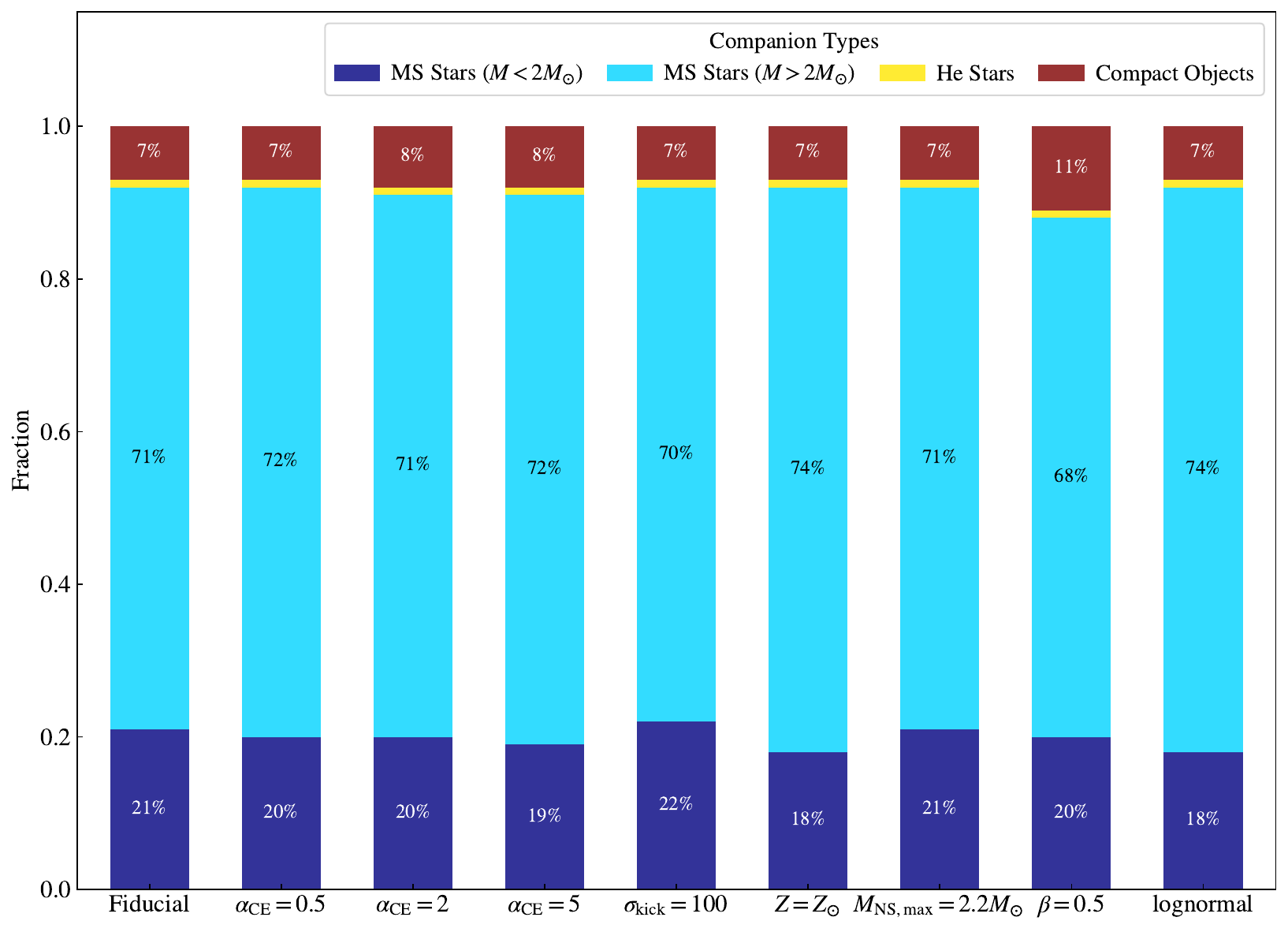}
\caption{Fractions of different magnetar companion types originating from disrupted binaries under different population synthesis models. The fraction of helium star companions is $\sim 1\%$.}
\label{fig:companion_fractions_disrp}
\end{figure*}

In addition to magnetars that remain in bound binary systems, a large fraction of magnetars are expected to form following the disruption of their progenitor binaries during the SN explosion (see section \ref{sec:fractions}). In such cases, the former companion star is ejected as a runaway object \citep[e.g.,][]{Tauris1998}, retaining valuable information about the evolutionary history of the magnetar progenitor.

Figure \ref{fig:companion_fractions_disrp} presents the distribution of magnetar companion types originating from disrupted binaries. We find that the majority of companions from disrupted systems are MS stars, predominantly with masses greater than $2\ M_{\odot}$. Only a negligible contribution arises from helium stars, which require envelope stripping through prior binary interactions. In addition, the fraction of compact object companions in the disrupted population is higher than that found among surviving magnetar binaries. These results suggest that runaway MS stars and compact objects represent the most promising targets for identifying the former companions of magnetars.

The companion counts are not weighted by the evolutionary lifetimes of the companions. This approximation is justified because the active magnetar phase is short \citep[$\sim 10^5$ yr, extending to at most $\sim 10^6$ yr in extreme cases;][]{Pons2009,Vigano2013,Rea2025}, and is shorter than the lifetimes of massive stellar companions. Consequently, neglecting companion lifetimes does not introduce a significant bias.

\subsubsection{Kick Velocities}

\begin{figure*}[htbp]
    \centering
    \includegraphics[width=0.85\textwidth]{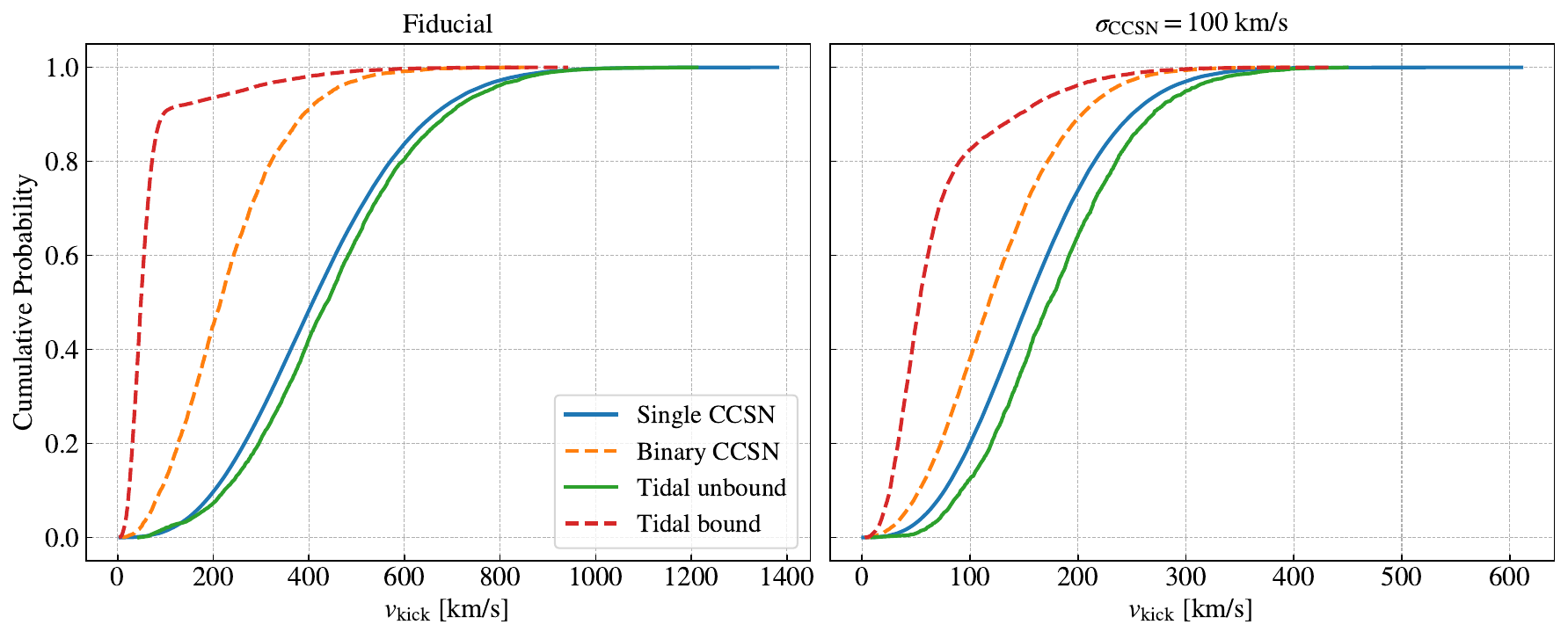}\vfill
    \includegraphics[width=0.85\textwidth]{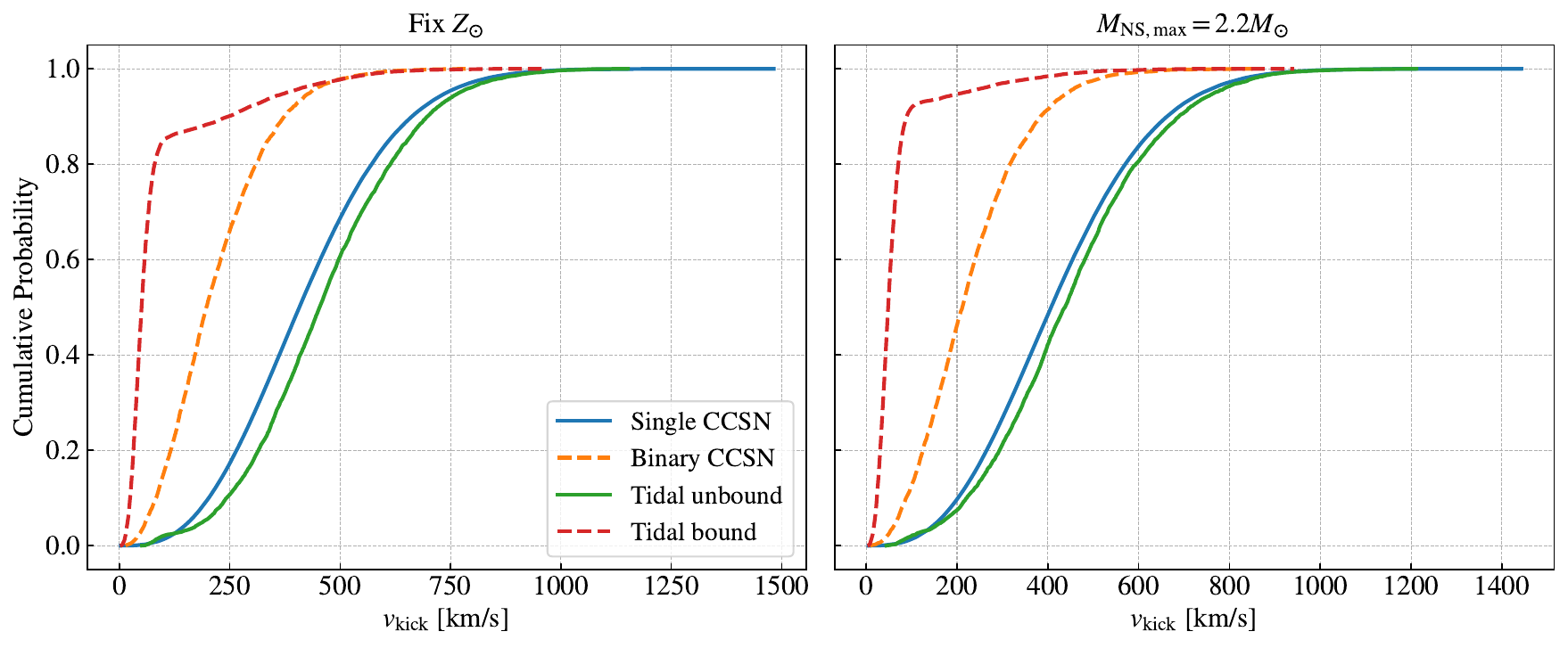}\vfill

    \caption{Cumulative distribution functions of magnetar kick velocities from different models (top left: fiducial; top right: $\sigma_{\mathrm{CCSN}}=100$ km/s; bottom left: fix solar metallicity; bottom right: $M_{\mathrm{NS,max}} = 2.2 M_{\odot}$). Each curve represents a different formation scenario, including single magnetars or magnetars in binary systems formed through CCSNe, as well as magnetars originating from tidal channels---specifically, those remaining bound in binaries ("tidal bound") and those ejected as single stars ("tidal unbound"). The x axis denotes the kick velocities, while the y axis shows the cumulative probability.}
    \label{fig:v_kick}
\end{figure*}

\begin{figure*}[htbp]
    \centering
    \includegraphics[width=0.95\textwidth]{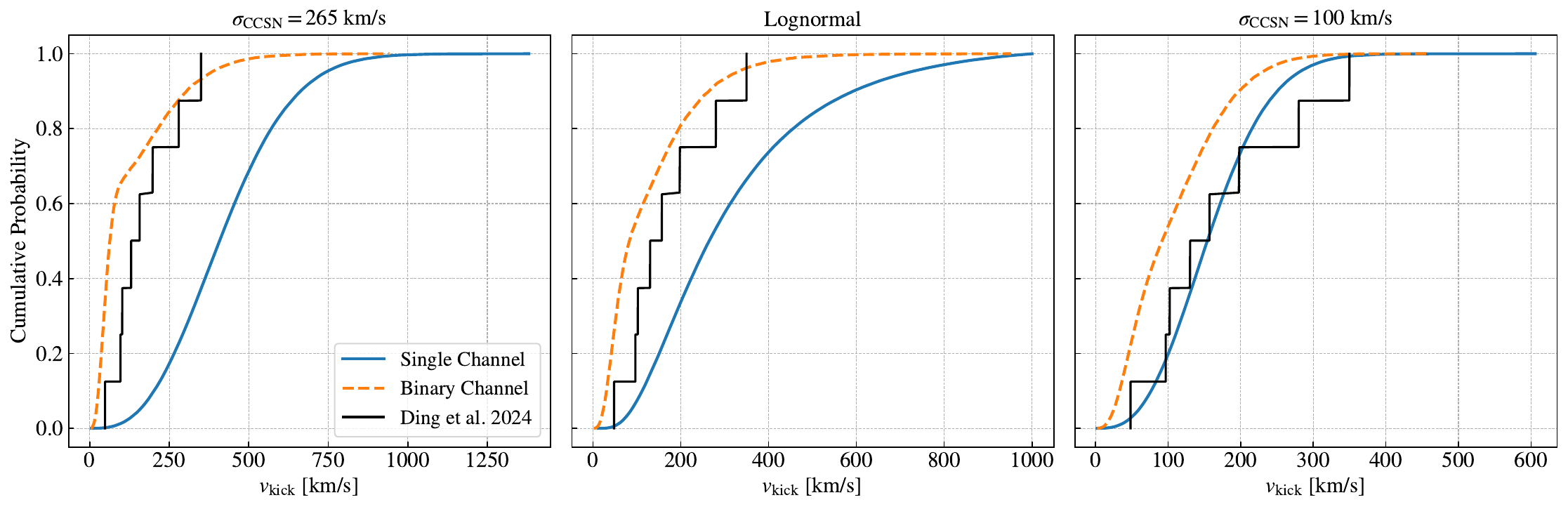}

    \caption{Same as the upper part of Fig.\ref{fig:v_kick}, but showing only the total distribution from channels forming single magnetars and channels forming magnetars in binaries. Black lines represent observational data from \cite{Ding2024}. }
    \label{fig:v_kick_com}
\end{figure*}

Figure \ref{fig:v_kick} presents the cumulative distribution functions of magnetar kick velocities across different evolutionary channels from different models. Channels involving CCSNe generally exhibit higher kicks, whereas those that remain in bound binary systems ("tidal bound") associated with ECSNe or USSNe yield smaller velocities. Compared to the fiducial model, adopting a smaller Maxwellian kick velocity dispersion shifts the entire distribution toward lower velocities, while assuming a fixed solar metallicity or a smaller maximum NS mass has little effect.

In Fig.\ref{fig:v_kick_com}, we combine all magnetar kicks into 2 categories, single channel and binary channel. In general, magnetars formed through single channels receive larger kicks, while those remaining bound in binaries have systematically lower kick velocities.
We compare these with the observed kick velocity distribution of Galactic magnetars \citep{Ding2024}, to assess which formation channels are most consistent with observations. It is worth noting that the observed kick velocities of Galactic magnetars are typically inferred from their transverse spatial velocities ($v_{\bot}$), which may not accurately represent their true three-dimensional natal kick velocities.
The results indicate that magnetars originating from single channels in the model with a Maxwellian kick velocity dispersion of $100$km/s provide the best agreement with the observed distribution of Galactic magnetars.
This suggests that Galactic magnetars tend to receive relatively small natal kicks at birth.

\subsubsection{Orbital Eccentricities and Periods}

\begin{figure*}[p]
    \centering
    \includegraphics[width=0.49\textwidth]{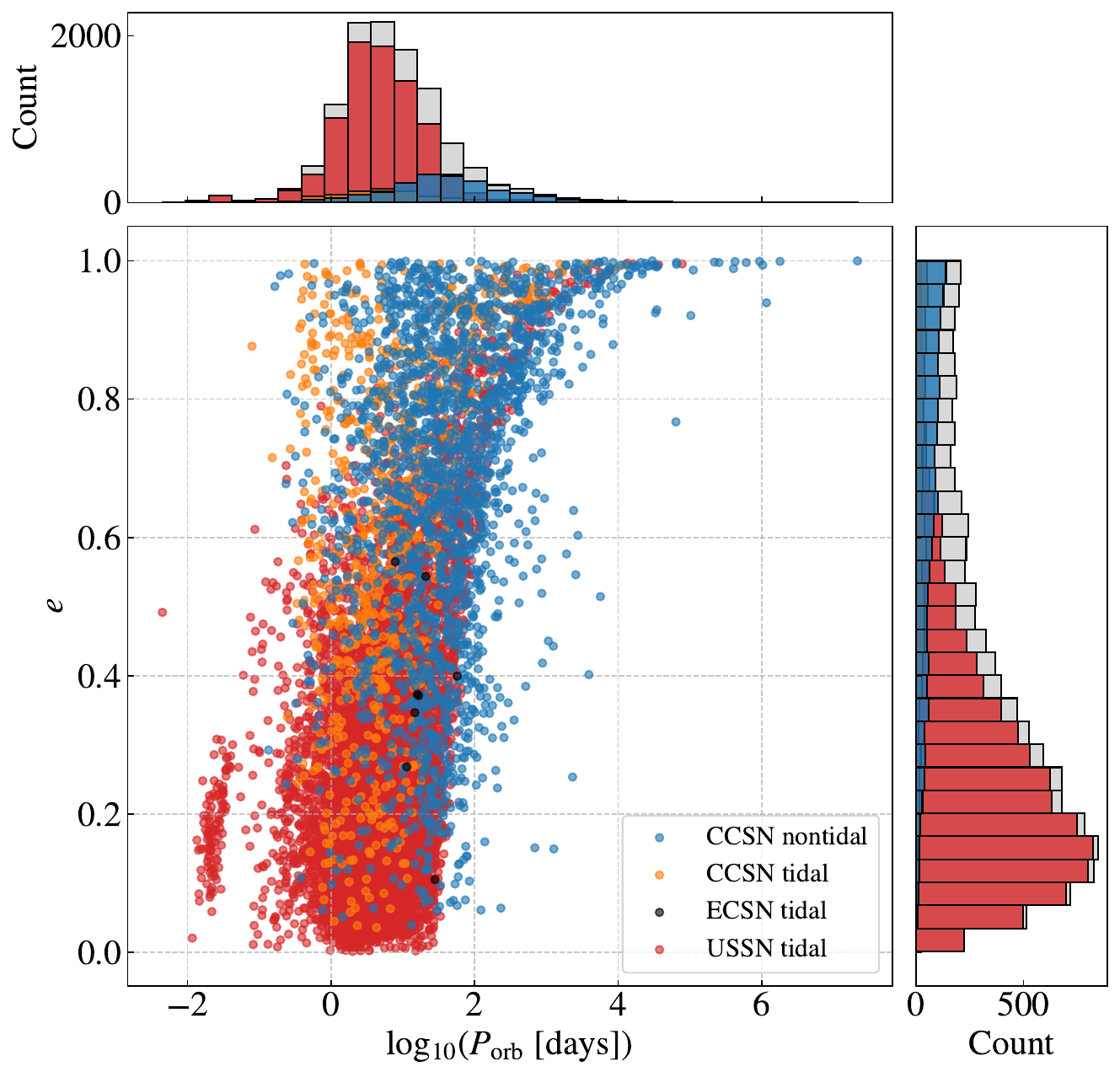}\hfill
    \includegraphics[width=0.49\textwidth]{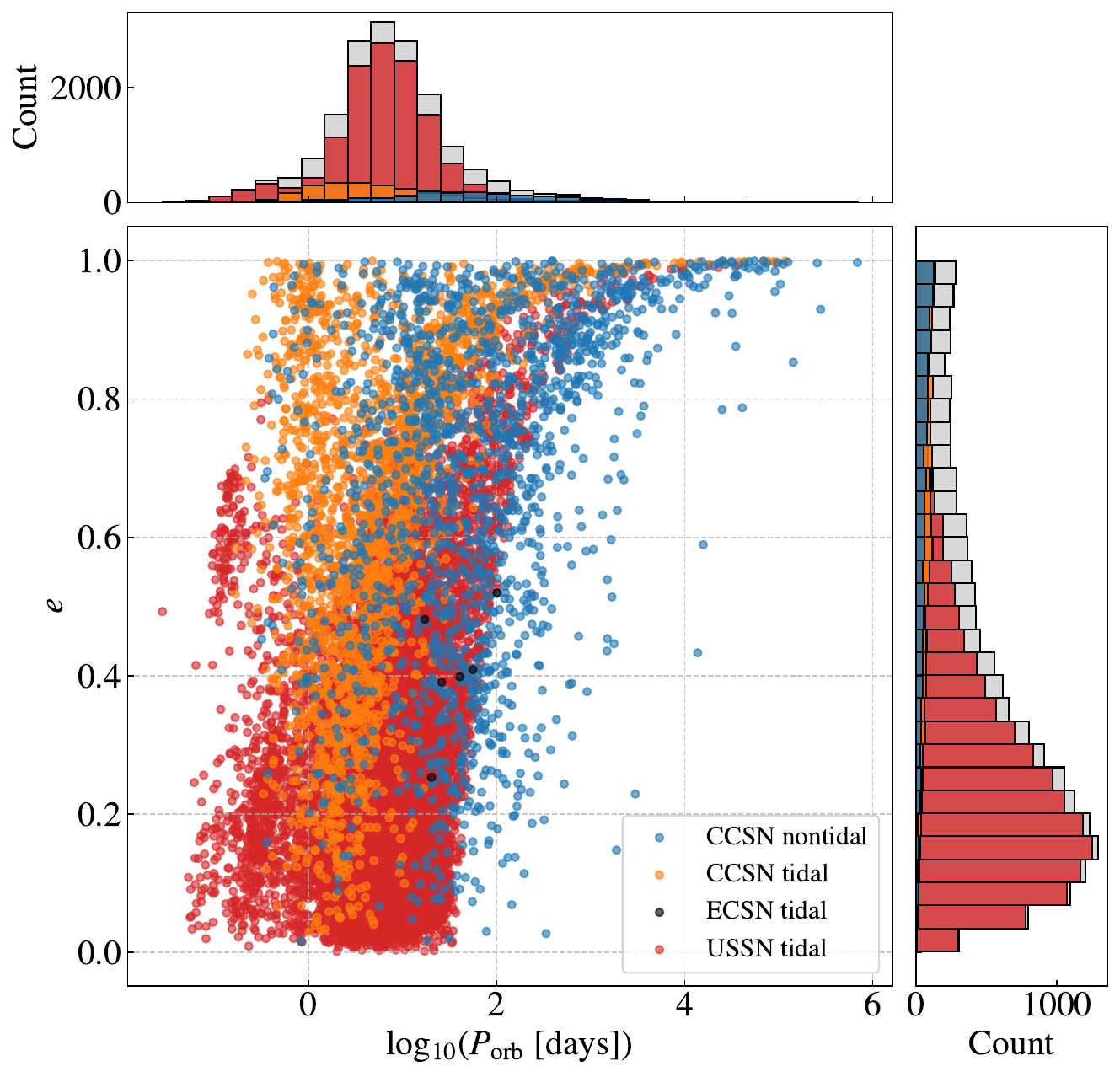}\hfill
    \includegraphics[width=0.49\textwidth]{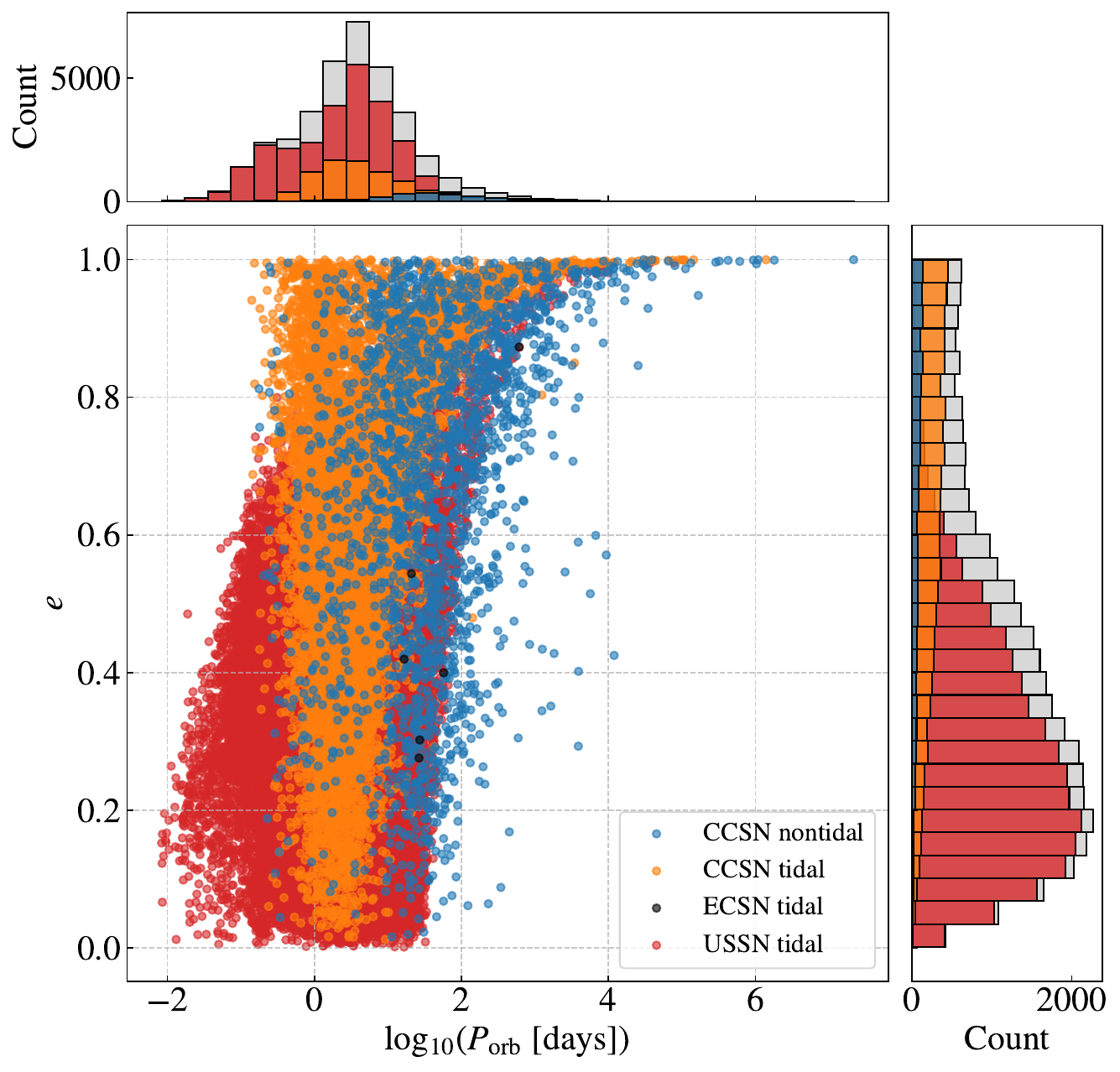}\hfill
    \includegraphics[width=0.49\textwidth]{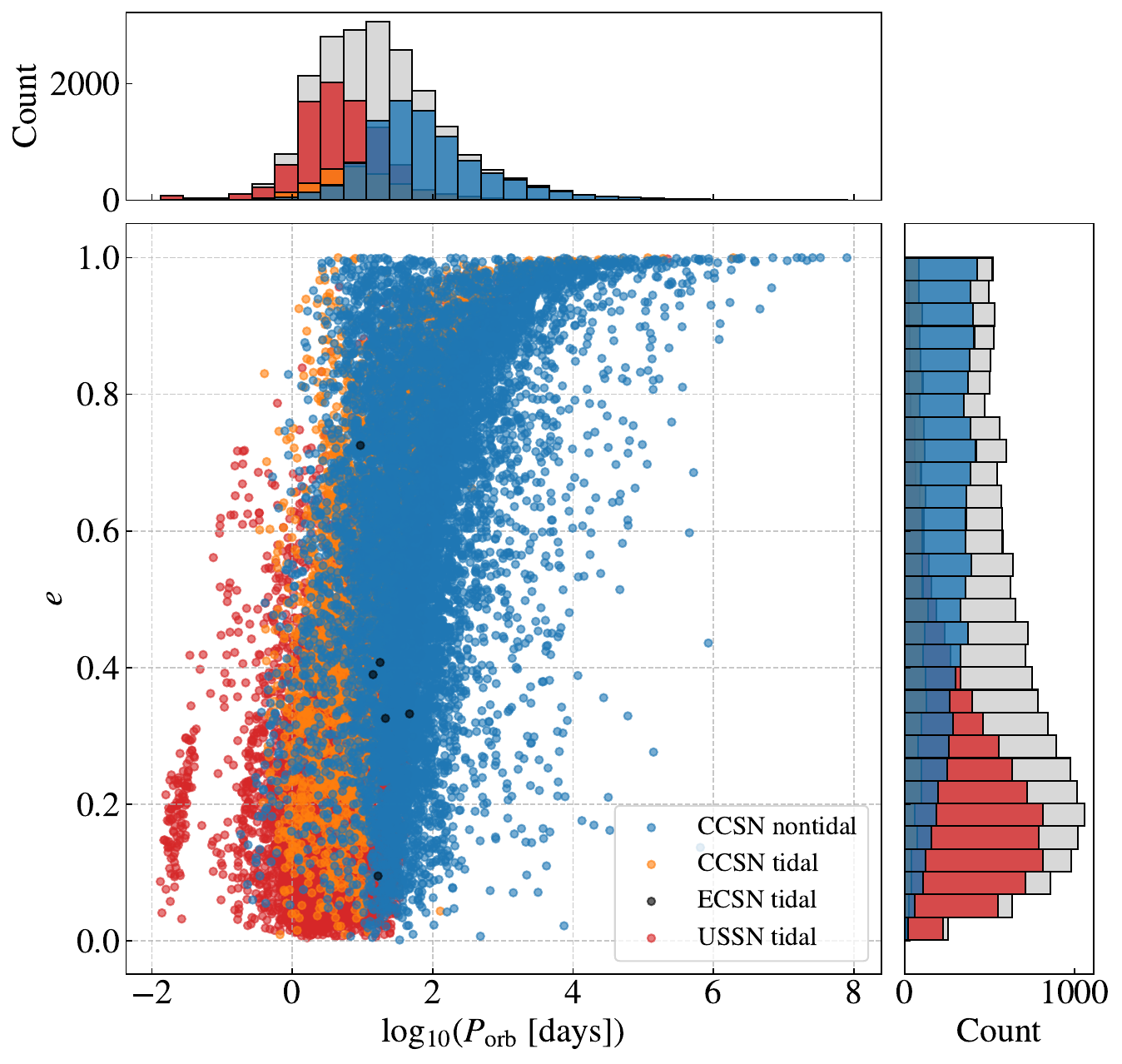}\hfill
    \caption{Distribution of orbital period and eccentricity for magnetar binary systems in different population synthesis models (top left: fiducial; top right: $Z=Z_{\odot}$; bottom left: $\alpha_{\mathrm{CE}}=5$; and bottom right: $\sigma_{\mathrm{CCSN}}=100$ km/s). Scatterplots show eccentricity vs. log orbital period for different SN types. Non-tidally affected CCSNe are shown in blue, while tidally affected CCSNe, ECSNe, and USSNe are shown in orange, light blue, and red, respectively. Top panel: the distribution of orbital period for each SN type and their combined total. Right panel: distribution of eccentricity for each SN type and total distribution.}
    \label{fig:e-p_compare}
\end{figure*}

The orbital period and eccentricity distributions of magnetar binary systems provide key insights into the effects of binary interactions and SN kicks on post-SN orbital dynamics. These parameters are particularly important for identifying potential observable binaries and understanding their formation pathways.

As presented in Fig. \ref{fig:e-p_compare}, in the fiducial model, which includes a metallicity-evolving population and standard SN kicks, most surviving magnetar binaries have orbital periods ranging from 0.1 to 100 days. The wide spread in eccentricity reflects the strong influence of SN kicks, which could impart significant asymmetry to the system. Magnetar binaries with shorter orbital periods tend to exhibit lower eccentricities and primarily originate from USSNe in the tidal spin-up channel, because of tighter pre-SN separations and weaker kicks. Magnetars formed through CCSNe typically receive stronger natal kicks, if the binary survives the explosion, the resulting system tends to have a larger orbital eccentricity.

Compared to the fiducial model, the fixed solar metallicity ($Z=Z_{\odot}$) scenario yields fewer systems with shorter orbital periods, consistent with stronger stellar winds at higher metallicity. However, the overall eccentricity distribution remains similar, indicating that metallicity has a minor effect on the post-SN orbital eccentricity compared to the SN kick velocity.
In the model with a larger $\alpha_{\text{CE}}$, a greater number of binaries survive the explosion, particularly those undergoing CCSN in the tidal spin-up channel. As a result, the surviving systems retain high eccentric orbits.
The most noticeable change occurs in the model with a reduced CCSN kick dispersion of $\sigma_{\mathrm{CCSN}} = 100$km/s. In this scenario, a greater number of wide-separation binaries that would typically be disrupted are able to survive the explosion, contributing to an increased population of systems with higher eccentricities and longer orbital periods.

These differences in period and eccentricity distributions may help distinguish between formation channels in future observations of magnetars in binaries. Systems formed with weaker natal kicks are more likely to remain in compact, low-eccentricity binaries.

\section{Discussion}
\label{sec:discussion}

\subsection{Implications for Magnetar-Related Objects}

The diversity of magnetar-related objects (e.g. GRBs, AXPs, FRBs, various types of SNe, and FBOTs) suggests that multiple formation channels may exit, imprinting different physical conditions on the newborn magnetars. Understanding these channels is essential to connect formation physics with observational phenomena.

Since the S2 channel contributes the most to the magnetar population, searching for the previous companions of Galactic magnetars provides a direct test of this formation pathway. However, while a runaway companion might in principle be detectable, Galactic magnetars are not newly formed, and any former companions may already have traveled far from the remnant site or evolved beyond easy recognition. 

Channels that efficiently spin up the stellar core, such as tidal interactions in close binaries, are promising ways to produce rapidly rotating magnetars. Such magnetars can deposit their rotational energy into the ejecta through spin-down and power stripped-envelope SNe \citep{Sautron2025}, such as Type Ic SLSNe, SNe Ic-BL, and a subset of FBOTs \citep{Liu2022,Hu2023}.

Although the majority of magnetars are predicted to be isolated, a small fraction survive in binary systems. Our simulations indicate that the most common surviving companions are MS stars, which can lead to magnetar X-ray binaries \citep{Xu2022}. More exotic configurations, such as helium star companions, are rarer but may explain a subset of ultraluminous X-ray sources.

Compact-object merger channels that produce magnetars may be probed via GWs, offering a direct test of their contribution to the magnetar population if the electromagnetic counterparts are detected.

\subsection{Uncertainties of Model Assumptions}

\subsubsection{Magnetar Formation through CCSNe}
\label{sec: uncertainties_CCSN}

In our model, the Tayler-Spruit dynamo is a key mechanism for magnetar formation. It is triggered when a PNS accretes sufficient material to generate strong differential rotation between its surface and interior. Therefore, its activation is related to the fallback accretion predicted by the CCSN models, which themselves carry significant uncertainties. 

\begin{figure}[t]
\centering
\includegraphics[height=9cm, trim = 0 0 0 0, clip]{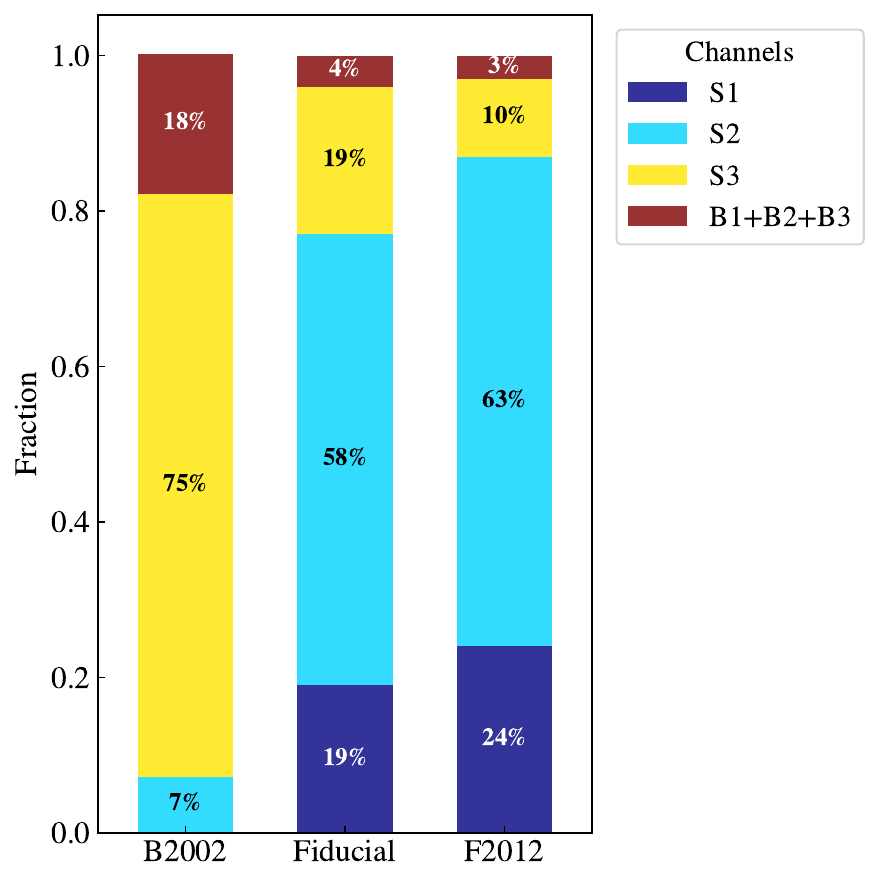}
\caption{Fractions of magnetars from different formation channels based on fallback accretion under different model assumptions: B2002, F2012, and our fiducial model.}
\label{fig:fration_CCSN_model}
\end{figure}

Different CCSN models predict varying amounts of fallback accretion onto the PNS. In \cite{Belczynski2002} (B2002), fallback accretion occurs only when the carbon-oxygen core mass exceeds $5 M_{\odot}$, resulting in negligible fallback in most cases. Consequently, only a small fraction of CCSNe experience sufficient fallback accretion to activate the Tayler–Spruit dynamo mechanism, leading to a very low fraction of magnetars among all NSs. In the \cite{Fryer2012} (F2012) delayed model, the fallback accretion mass exceeds $0.2 M_{\odot}$, implying that all CCSNe would generate magnetars, leading to a very large fraction of magnetars among all NSs. 

Figure \ref{fig:fration_CCSN_model} shows the fractions of magnetars formed through different evolutionary channels based on B2002, F2012 and, as comparison, our fiducial model. However, both scenarios appear to be inconsistent with the current constraints on the magnetar fraction among NSs. Given the significant uncertainties in CCSN modeling, we instead use the fraction of magnetars among all NSs to constrain the fraction of CCSNe that trigger the Tayler–Spruit dynamo and form magnetars.

\begin{figure*}[t]
\centering
\includegraphics[height=9cm, trim = 0 0 0 0, clip]{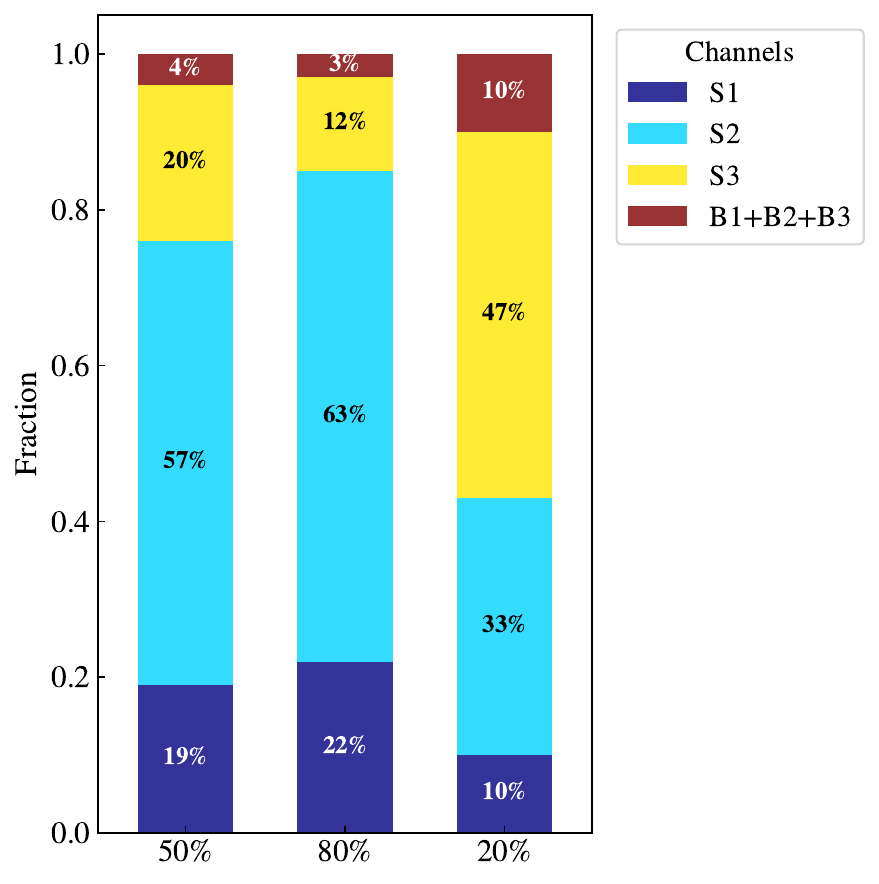}
\hspace{-0.2cm}
\includegraphics[height=9cm, trim = 0 0 135 0, clip]{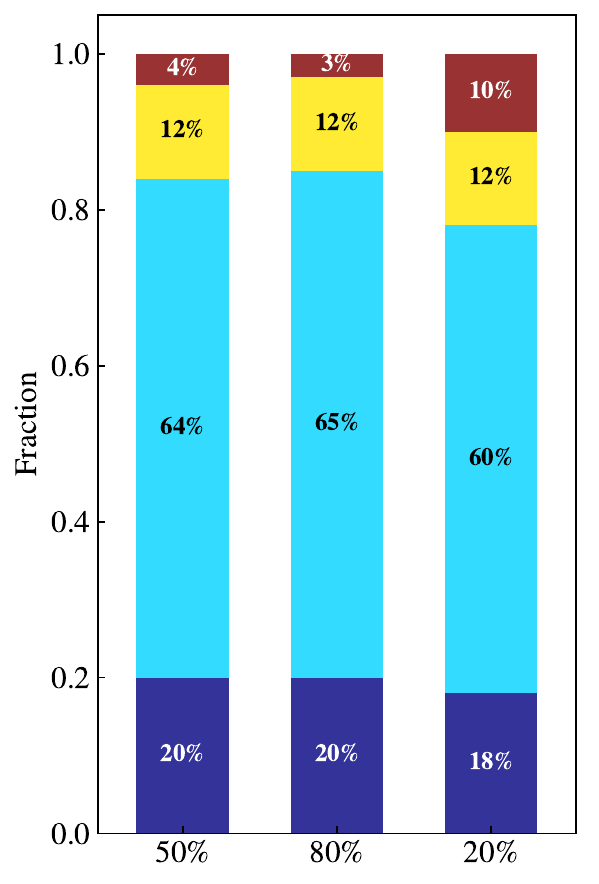}
\caption{Fractions of magnetars from different formation channels based on the fiducial model. Left panel: all BMS mergers lead to magnetar formations. Right panel: massive stars form BMS mergers only produce magnetars through triggering Tyler-Spruit dynamo. Different bars represent different magnetar formation fractions among all NSs.}
\label{fig:fration_TS_mag}
\end{figure*}

As presented in the left panel of Fig.\ref{fig:fration_TS_mag}, we use different assumed magnetar formation fractions (20\%, 50\%, 80\%) to constrain the expected contribution from fallback-induced magnetars.
As shown in the figure, increasing the assumed magnetar formation fraction leads to a higher contribution from fallback-induced magnetars, particularly within the S1 and S2 categories. Conversely, when the assumed magnetar formation fraction is low, the contribution from CCSNe becomes small, and the merger channel becomes the dominant one for magnetar formation.

\subsubsection{Magnetars from BMS Merger Channel}
\label{sec: uncertainties_BMS}

Another uncertainty lies in the assumption that all BMS mergers produce massive MS stars with strong magnetic fields and rapid rotation. Observationally, approximately 7\% of massive stars exhibit strong surface magnetic fields—a fraction that aligns closely with the proportion of massive stars in our simulation that have undergone stellar mergers. This agreement supports the hypothesis that stellar mergers may be the primary origin of magnetic fields in massive stars. In the right panel of Fig.\ref{fig:fration_TS_mag}, we also present an alternative scenario representing a lower limit on the contribution from the merger channel. In this case, no massive stars possess strong magnetic fields from stellar mergers. They can only form magnetars via the Tayler–Spruit dynamo during CCSNe.

\subsubsection{Population Synthesis Variations}

Observational surveys indicate that a large fraction of massive stars reside in binary or higher-order multiple systems \citep{Sana2012,Sana2014,Moe2017}. Although we include binaries in our simulations, higher-order multiples (triples or quadruples) are not modeled, which could open additional pathways for magnetar formation. The precise fraction of massive stars in such systems, and their orbital configurations, remain uncertain and may influence the predicted channel fractions.

Our models vary key parameters such as the CE efficiency ($\alpha$), natal kick distribution, and maximum NS mass individually to assess their impact. While informative, this one-parameter-at-a-time approach does not capture possible correlations between parameters or the full complexity of the model space. More comprehensive approaches, such as statistical sampling of multidimensional parameter grids \citep{Broekgaarden2021} or Bayesian inference frameworks \citep[e.g.,][]{Barrett2018,Wong2023}, could provide a smoother exploration and help identify parameter combinations that best reproduce observed magnetar and compact-object populations.

Additional assumptions include the distributions of initial binary mass ratios and orbital periods, tidal interaction efficiencies, and the detailed modeling of magnetic-field amplification. Each introduces additional uncertainty in the predicted magnetar population, though they are difficult to quantify precisely in rapid population synthesis codes.

\subsection{Comparison with Previous Population Synthesis Studies}

Previous population synthesis studies of magnetars have primarily focused on the statistical properties and postbirth evolution of isolated NSs, incorporating magnetothermal evolution models to demonstrate how magnetic-field decay and spin-down shape the observed distributions in the $P$–$\dot{P}$ plane and to explore possible evolutionary links among different NS classes \citep[e.g.,][]{Popov2002,Popov2010,Vigano2013,Gullon2014,Beniamini2019,Sautron2025}. These studies have been successful in reproducing the observed properties of magnetars. However, they did not explicitly address the formation channels of magnetars or the role of binary interactions in determining their birth properties.

Our work focuses on the formation channels of magnetars by simulating the evolution of massive stars and binary systems prior to core collapse. By linking magnetar birth properties to specific progenitor pathways and binary interactions, our results complement earlier population synthesis studies, providing a physical basis for the initial conditions adopted in postbirth evolutionary models of the magnetar population.

\section{Conclusion}
\label{sec:conclusion}

Magnetars are linked to a range of high-energy astrophysical transients, yet their formation channels and population properties remain uncertain.
In this paper, we used population synthesis simulations to explore the contribution of various formation channels to the magnetar population, including single-star evolution, and isolated binary evolution. Our main conclusions are as follows.

\begin{itemize}
    \item Our simulations indicate that, while the majority of magnetars are observed as isolated NSs, a significant fraction of them originate from binary interactions, underscoring the critical role of binary evolution in shaping the magnetar population.
    \item Among the various formation channels, magnetars from the S2 channel (kick-disrupted binaries) contribute the most to the total magnetar population in our fiducial model, accounting for approximately 57\% of all magnetars. The dynamo mechanism operating in differentially rotating PNSs is primarily responsible for amplifying magnetic fields to magnetar strength in this channel.
    \item The delay time between progenitor formation and magnetar birth depends strongly on the evolutionary channel. While single channels contribute mainly to prompt magnetar formation, binary channels broaden the distribution, producing both very short delays (e.g., tidal spin-up) and extended tails through processes such as AIC. As a result, magnetars can arise across a wide range of stellar ages, from young star-forming environments to evolved populations, with binary evolution playing a central role in shaping this diversity.
    \item Although most binaries are disrupted by SN kicks, our simulations show that a small fraction of magnetars survive in binary systems. Those surviving binaries are most likely to host MS companions, reflecting the tendency of binaries to remain bound when the mass ratios are favorable and the natal kicks are moderate. 
    \item Our simulations reveal that the low–dispersion CCSN kick model produces single-channel magnetars whose velocity distribution agrees most closely with that inferred from Galactic magnetar observations. This consistency suggests that the Galactic magnetar population may preferentially originate from progenitors that receive weaker natal kicks than typically assumed for NSs. Such a trend could reflect the influence of ultrastripped or fallback–dominated SNe, or point toward systematic differences between magnetar and ordinary NS formation.
    \item Our simulations reveal that the eccentricity–period distribution is shaped by the interplay between progenitor metallicity, CE efficiency, and the specific SN mechanism.
    In the fiducial model, most surviving systems have periods spanning from 0.1 to 100 days with a broad eccentricity spread, where short-period, low-eccentricity binaries arise mainly from USSNe in the B2 channel, while CCSNe contribute to highly eccentric systems. Comparisons across models indicate that metallicity chiefly influences the orbital period distribution, with higher-metallicity populations yielding fewer short-period systems due to stronger stellar winds. In contrast, variations in CE efficiency and natal kick strength have a more pronounced effect: a higher CE efficiency increases the survival of systems from CCSNe, while reduced CCSN kicks allow wider and more eccentric binaries to persist. 
\end{itemize}

Upcoming FRB and GW surveys may help constrain the contribution of binary and merger channels to the magnetar population. Further improvements to fallback accretion modeling will refine predictions of magnetar formation from CCSNe.

\begin{acknowledgments}
We thank the anonymous referee for helpful suggestions and Jeffrey Andrews, Abhishek Chattaraj, Hailiang Chen, Xuefei Chen, Zhanwen Han, Nanda Rea, Myles Sherman and Yuan-Pei Yang for useful discussions or comments. We acknowledge the Nevada Center for Astrophysics, and a UNLV Top-Tier Doctoral Graduate Research Assistantship (TTDGRA) for support. This work was supported by computational resources provided by Expanse \citep{strande2021expanse}.
\end{acknowledgments}

\vspace{5mm}

\software{\texttt{COMPAS} {\citep[version 03.12.00;][]{Stevenson2017,Vigna2018,Neijssel2019,TeamCOMPAS2022,TeamCOMPAS2025}};  \texttt{Python}, \url{https://www.python.org}.
          }

\appendix

\section{Population Synthesis Settings}
\label{sec: app_setup}

Table \ref{tab:COMPAS} summarizes the initial values and parameter settings adopted in our population synthesis simulations using {\texttt{COMPAS}}.

\begin{table*}
\caption{{Initial values and settings of the population synthesis simulation with {\texttt{COMPAS}}  }}
\label{tab:COMPAS}
\centering
\resizebox{\textwidth}{!}{\begin{tabular}{lll}
\hline  \hline
Description and name & Value/range & Note/setting   \\ 
\hline  \hline
\multicolumn{3}{c}{Initial conditions} \\ 
\hline
Initial primary mass $M_{1,\rm i}$ & $[5, 150]\,M_\odot$  & \citet{Kroupa2001} IMF  $\propto  {M_{1,\rm i}}^{-\alpha_{\rm IMF}}$  with $\alpha_{\rm{IMF}} = 2.3$ for stars above $5\,M_\odot$	  \\
Initial mass ratio $q_{\rm i} = M_{2,\rm i} /  M_{1,\rm i} $ & $[0, 1]$ & We assume a flat mass ratio distribution  $p(q_{\rm i}) \propto  1$ with $M_{2,\rm i}\geq 0.1\,M_\odot$  \\
Initial semi-major axis $a_{\rm i}$ & $[0.01, 1000]\,{\rm AU}$ & Distributed flat-in-$\log p(a_{\rm i}) \propto 1 / {a_{\rm i}}$ \\   
Initial metallicity $Z_{\rm i}$ & $[0.0001, 0.03]$ & Distributed using a uniform grid in $\log (Z_{\rm i})$         \\
Initial orbital eccentricity $e_{\rm i}$ & 0 & All binaries are assumed to be circular at birth  \\
\hline
\multicolumn{3}{c}{Parameter settings:} \\ 
\hline
Stellar winds for helium stars &  \cite{Sander2020} & Based on   \cite{Sander2020} and temperature correction from \cite{Sander2023}  \\
Max transfer stability criteria & $\zeta$-prescription & Based on \cite{Vigna2018} and references therein     \\ 
Mass transfer accretion rate & thermal timescale & Limited by thermal timescale for stars  \cite{Vink2005,Vinciguerra2020} \\ 
 & Eddington-limited  & Accretion rate is Eddington-limit for compact objects  \\
Non-conservative mass loss & isotropic re-emission &  {\citet[][]{Massevitch1975,Bhattacharya1991}}; \\ 
& &  {\cite{Soberman1997,Tauris2006}} \\
Case BB mass transfer stability & always stable & Based on  \citet{Tauris2015,Tauris2017,Vigna2018} \\ 
CE prescription & $\alpha_{\rm CE}-\lambda$ & Based on  \citet{Webbink1984,DeKool1990}  \\
$\alpha_{\rm CE}$-parameter & 0.5, 1, 2, 5 &  \\
CE $\lambda$-parameter & $\lambda$ & Based on \cite{Xu2010} and \cite{Dominik2012}  \\
Hertzsprung gap (HG) donor in {CE} & pessimistic & Defined in \citet{Dominik2012}: HG donors don't survive a {CE} phase \\
{SN} natal kick magnitude $v_{\rm k}$ & $[0, \infty)\,{\rm km}\,{\rm s}^{-1}$ & Drawn from Maxwellian distribution with standard deviation $\sigma_{\rm{rms}}$          \\
{SN} natal kick polar angle $\theta_{\rm k}$ & $[0, \pi]$ & $p(\theta_{\rm k}) = \sin(\theta_{\rm k})/2$ \\
{SN} natal kick azimuthal angle $\phi_{\rm k}$ & $[0, 2\pi]$ & Uniform $p(\phi) = 1/2\pi$   \\
{SN} mean anomaly of the orbit & $[0, 2\pi]$ & Uniformly distributed  \\
CCSN remnant mass prescription  & delayed &  From \citet{Fryer2012}, which  has no lower {BH} mass gap  \\
USSN remnant mass prescription & delayed &  From \cite{Fryer2012} \\
ECSN  remnant mass presciption & $m_{\rm{f}} = 1.26\,M_\odot$ & Based on Equation (8) in \citet{Timmes1996}          \\
CCSN velocity dispersion $\sigma_{\rm{rms}}$ & $100,\,265\,{\rm km}\,{\rm s}^{-1}$ & 1D rms value based on \cite{Hobbs2005} and \cite{Atri2019} \\
USSN and ECSN velocity dispersion $\sigma_{\rm{rms}}$ & $30\,{\rm km}\,{\rm s}^{-1}$ & 1D rms value based on e.g., \cite{Pfahl2002,Podsiadlowski2004}    \\
CCSN natal kick distribution & lognormal & $\mu=5.6\, , \sigma=0.68$ based on \cite{Disberg2025}\\
PISN/PPISN remnant mass prescription & \cite{Marchant2019} & As implemented in \cite{Marchant2019}      \\
Maximum NS mass & $M_{\rm NS,max}=2.2,\,2.5\,M_\odot$ & Based on \cite{Antoniadis2013,Alsing2018,Romani2022}            \\
Tides and rotation & & We do not include prescriptions for tides and/or rotation\\
\hline
\multicolumn{3}{c}{Simulation settings} \\ 
\hline
Total number of systems sampled per model  & $10^6$ & \\  
Solar metallicity $Z_\odot$ & $Z_\odot$ = 0.0142 & Based on {\cite{Asplund2009}} \\
Binary population synthesis code                                      & \texttt{COMPAS} (v03.12.00) &  \cite{Stevenson2017,Vigna2018,Neijssel2019}; \\
& & \cite{Broekgaarden2019,TeamCOMPAS2022}; \\
& & \cite{TeamCOMPAS2025}\\
\hline \hline
\end{tabular}
}
\end{table*}

\bibliography{main}{}
\bibliographystyle{aasjournal}

\end{document}